\documentclass[twocolumn,showpacs,aps,pre]{revtex4}
\usepackage{graphicx}
\begin{document}


\title{Radial and axial segregation in horizontal rotating cylinders studied by Magnetic Resonance Imaging (MRI)}

\author{Thoa T.\ M.\ Nguy$\tilde{\hat{\mbox{e}}}$n}
	\email{ttmn2@cam.ac.uk}
\author{Andrew J.\ Sederman}
\author{Mick D.\ Mantle}
\author{Lynn F.\ Gladden}
\affiliation{Department of Chemical Engineering, University of Cambridge, Cambridge CB2 3RA, United Kingdom.}

\date{\today}

\begin{abstract}
The dynamics of granular materials, mostly radial and axial segregation in horizontal rotating cylinders filled with millet and poppy seeds, is studied by Magnetic Resonance Imaging (MRI). For the first time, full 3D structures and real-time 2D MRI movies showing the progress of segregation over many hours are reported. Data are acquired with sufficiently high temporal and in-plane spatial resolutions (74 ms and 0.94 mm $\times$ 0.94 mm, respectively), giving new insights into the underlying mechanisms. The millet and poppy mixture composition is calibrated based on the signal intensity and is quantified throughout segregation. As for radial segregation, millet and poppy mixture core formation is observed in cylinders of 75\% and 82\% filling level. The size of the core is calculated and the avalanche layer thickness is therefore determined. 2D MRI movies showing real-time radial segregation suggests that initial radial segregation is due to the stopping and percolation of poppy seeds. Axial segregation is characterized by the formation, traveling and merging of poppy-rich bands. In all cases studied, the formation of poppy-rich bands is observed, after which individual bands start to travel at $\sim$3 $\mu$m s$^{-1}$ until they are within $\sim$3 cm of a stationary band. Adjacent bands then merge into a single, enlarged poppy band as millet seeds move out of the merging region. In both radial and axial segregation, core diffusion is shown to be another mechanism pathway for segregation besides the free surface.
\end{abstract}

\maketitle

\section{\label{sec:level1-1}Introduction}
Granular materials, including grains, gravels, and many powders, are aggregates of large solid particles. They are known to have a highly complex disordered structure, nonlinear internal friction, tend to form stable arches and voids, and inelastic collision between particles under load \cite{painter97}. As a consequence, granular materials possess many physical properties which are not observed in classical solid or liquid state physics, such as internal stress fluctuations \cite{gioia06}, intermittent avalanches \cite{rajchenbach97}, segregation \cite{ottino00}, and spatial pattern creation \cite{hill05}. 

Horizontal rotating drums are common in industry, such as tumbling mixers and reactors, and the segregation of granular materials in this geometry is of much interest. In such systems, segregation can occur radially and axially. This is a well-known phenomenon and has been reported many times in the literature. However, there are still areas of controversies, such as the mechanism of radial segregation, the role of core diffusion versus that of the free surface, which will be reviewed in turn below. In addition, this work also reports new observations of radial segregation and band traveling and merging. This area has only been touched upon in the literature, so a review of the current knowledge is also included.

The mechanism that drives radial segregation has been considered in the literature. There is an argument that radial segregation is due to competing attractors, one near the outer edge that has higher affinity for larger particles and one near the rotation axis that has higher affinity for smaller particles \cite{clement95}. However, a different approach is based on the statistical mechanics in which the granular material is treated as a sample of many small particles and their collective behavior depends on the sample volume and compactivity \cite{edwards89}. Another explanation attributes this to the percolation and stopping of small particles \cite{cantelaube95}. It is reasoned that as a mixture of particles flow down the slope in the avalanche layer, the smaller particles carry less momentum due to their lower mass and therefore are more likely to stop in the middle of the slope during collisions between particles while the larger still keep rolling downward and get into the outer layer. This creates segregated layers of smaller particles near the bottom of the flowing layer and larger near the top \cite{hill05}. As a result, smaller particles do not tend to reach the cylinder wall at the bottom of the inclined avalanche region but instead form a core along the rotation axis. This process is very rapid, which renders radial segregation visible within one revolution and possibly complete within a few. 

The free surface has mostly been thought to play the sole role in both radial and axial segregation so far, especially in modeling work. Lai et al.\ simulated radial segregation due to friction by using the BTW (Bak, Tang, and Wiesenfeld) sand-pile model \cite{lai97}. However, the rotating cylinder was assumed to mimic a sand pile, so no consideration was given to the geometry of the cylinder, nor its rotational speed. These details were, however, included in a later and more detailed study, together with the dynamics of the particles in the thin avalanche layer \cite{khakhar97}. This was based on the density difference of the same sized-particles and the mechanism of the segregation was assumed to be purely due to the avalanche layer. As will be shown later, this is only valid for systems of less than 75\% filling. Other 2D models have also been considered but restricted to either single particle dynamics \cite{clement95}, or low rotational speed \cite{mccarthy96}. In all cases, the avalanche layer was considered to be the only mechanism for segregation. Other authors propose that diffusion inside the core may play a role \cite{hill97}. From geometrical considerations, it is evident that all seeds in the bulk reach, and are therefore influenced by, the avalanche layer in every rotation in a cylinder less than 50\% full. However, at higher filling fraction, there is a cylindrical core about the rotation axis which may never reach the avalanche layer. It is in our interest to study such systems to examine the underlying mechanisms.

Upon radial segregation, further rotations lead to the formation of axial bands at equal spacings, as reported by some authors to date \cite{hill97}. However, the evolution of bands does not stop at band formation but develops into band traveling and merging over a longer time scale. In one study, the particle velocity in the flowing layer was found to be higher for larger particles, and thus thought to be the main drive for axial segregation \cite{newey04} using the camera technique. This method involves the imaging of the cylinder surface by a charged coupled device (CCD) camera and the velocity is calculated by tracking individual particles. The data were then, together with the frictional properties, successfully implemented into a Monte Carlo simulation to reproduce band merging. However, this model is mainly qualitative, segregation is modelled to occur much faster than reality and has no prediction of what will happen beyond the first merging. Using Magnetic Resonance Imaging and camera techniques, other authors also reported band merging \cite{choo98,hill97}. The MRI data give some qualitative insight into the progress of the merging bands but only in terms of visual observation and is not sufficient as only 2D stationary images were acquired. A continuum model based on two local variables, concentration difference and the dynamic angle of repose, was subsequently proposed to fit the data obtained by the camera technique \cite{aranson99}. However, the model was limited to predicting when the merging would occur, without revealing what would happen to the particles at the front of and between the merging bands. Nonetheless, the author was successful at working out a logarithmic rate of reduction of the number of bands over time, which is again confirmed in a later study \cite{fiedor03}. No real 3D data are available to give full description of the mechanisms of band merging. 

In this study, radial and axial segregation are investigated using MRI, which is very useful for the study of granular flows as most of the systems are opaque. Dyed beads have been used but only surface observations can be made and are inappropriate for examining the state of segregation inside the bulk material. This problem can be rectified by using particles with different MRI activities, which renders them distinguishable when mapped in space and time. The theory of MRI can be found in an excellent text by Ref.\ \cite{callaghan91}. 

\begin{figure}
\includegraphics[width=0.7\columnwidth]{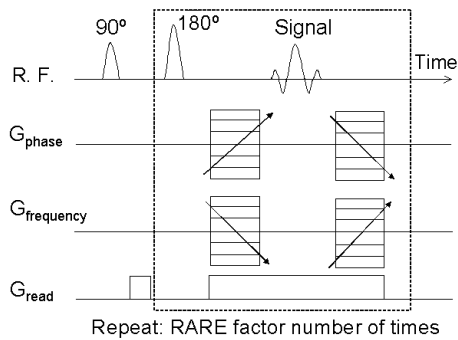}
\caption{\label{fig1} 3D RARE pulse sequence diagram. Homospoil can be used before and after the $90^\circ$ pulse.}
\end{figure}

The main pulse sequence used in this work is 3D RARE (Rapid Acquisition with Relaxation Enhancement) with the pulse diagram shown in Fig.\ \ref{fig1} \cite{hennig86}. This is a fast imaging technique that samples multiple-line k-space from each excitation, thus reducing the acquisition time.

\begin{figure}
\includegraphics[width=0.7\columnwidth]{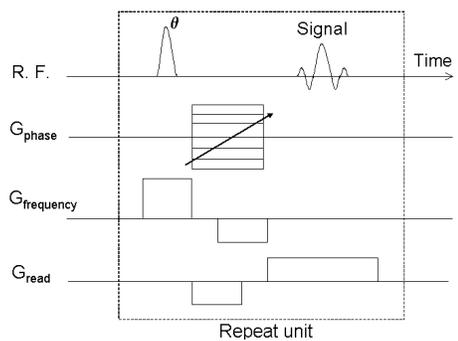}
\caption{\label{fig2} 2D FLASH pulse sequence diagram. The readout gradient is often prolonged to act as homospoil.}
\end{figure}

The disadvantages of the RARE sequence, though having good signal to noise ratio (SNR), are that it cannot image moving objects and its acquisition time is of the order of minutes for 3D or many seconds for 2D images, therefore not sufficiently rapid to acquire images in real time. For this reason, 2D FLASH (Fast Low Angle SHot) is instead employed, the pulse sequence of which is shown in Fig.\ \ref{fig2} \cite{haase86}. The main feature of FLASH is the very low tip angle $\theta$ used, typically $5^\circ-10^\circ$, and thus has an inherently low SNR \cite{mantle03}.

In this work, full 3D structures and real-time 2D movies are presented, providing new insights into segregation phenomena. New observations of radial and axial segregation are made and issues brought up in the literature review are addressed, by conducting a number of experiments listed below:

\renewcommand{\theenumi}{\alph{enumi}}
\begin{enumerate}
\item	New observations: core formation throughout radial segregation in cylinders of 75\% and 82\% full cylinders. The avalanche layer thickness can then be inferred from the core size. Cylinder length is 15 cm and 12 cm, respectively, but has no effect on radial segregation.
\item	Mechanism of radial segregation: real-time 2D movies showing the stopping and percolation mechanism. Experiments are conducted with an 82\% cylinder, rotated at 10 rpm. Radial segregation occurs slowly in such a full cylinder at a low rotational speed, therefore assisting data acquisition.
\item	The role of core diffusion versus that of the free surface: can be seen in both radial and axial segregation and is examined in both sections \ref{sec:level2-1} and \ref{sec:level2-2}. Supporting data are acquired during radial segregation in cylinders of 75\% and 82\% full; as well as during axial segregation in an 82\% full cylinder. Only cylinders of more than 75\% filling are chosen because the effect of the free surface movement and core diffusion can be uncoupled in a core buried deep under the avalanche layer.
\item	New observations: band traveling and merging. The long time scale 3D RARE data are acquired for a 50\% full and 50 cm long cylinder, rotated at 35 rpm. The short time scale 2D FLASH data are acquired for a 50\% full cylinder, also rotated at 35 rpm, with the lenth of 12 cm to fit within the FOV. Only 50\% full cylinder is examined because axial dynamics occurs much more rapidly compared to that in 75\% or 82\% full cylinders.
\end{enumerate}

\section{\label{sec:level1-2}Methodology}
The system to be studied is a closed horizontal cylinder of 48 mm I.D., filled with a mixture of poppy and millet seeds. Both seeds are MRI active due to their oil content but poppy seeds give a signal ($5000\ \mbox{au}\ \mbox{pixel}^{-1}$) 10 times stronger than millet ($500\ \mbox{au}\ \mbox{pixel}^{-1}$). Poppy seeds are generally ellipsoidal with a major axis of 1 mm and minor axes of 0.7 mm, approximately; whereas millet seeds are more or less spherical of 2.5 mm in diameter. Poppy and millet have similar density (1100 kg m$^{-3}$) so their segregation is expected to be mainly size-driven \cite{kakalios05}. It was found from previous experiments that segregation was most obvious when the poppy seeds were less than millet in volume so a composition of 25\%:75\% vol (poppy : millet) was adopted for all  experiments. The angle of repose of poppy seeds is approximately $2^\circ$ higher than that of millet, which is just over $30^\circ$.

The cylinder length is adjustable and chosen depending on the application. A typical length is 12 cm so as to fit in the FOV of the magnet. However, a 50 cm length has been used so that band traveling and merging can be observed. The cylinder filling fraction is also varied depending on the purpose of the experiment. 75\% and 80\% full cylinders are used for experiments studying the role of core diffusion versus that of the free surface as these two effects can be uncoupled. 50\% is chosen to study axial segregation as it is conventional and occurs much more rapidly to assist data acquisition.

\begin{figure}
\includegraphics[width=\columnwidth]{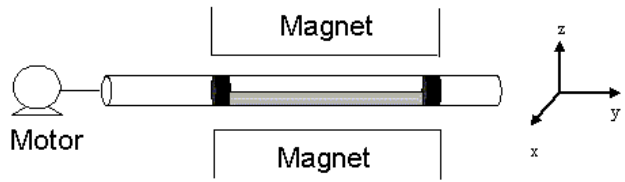}
\caption{\label{fig3} Schematic diagram of the experimental setup and its coordinates.}
\end{figure}

The seeds are premixed externally before loaded into the Perspex tube, which is rotated by an electric motor located about 2 m away from the horizontal Bruker Biospec AV magnet (85 MHz, $^1H$-resonance, 2 T) as shown in Fig.\ \ref{fig3}. The two adjustable dividing plugs allow variable tube lengths of interest.

All 3D MRI data are acquired using a RARE sequence while the motor is at rest. The RARE pulse sequence is implemented as follows: The $90^\circ$ and $180^\circ$ soft pulses are both 512 ms sinc pulses. The spin-lattice relaxation time $T_1$ is measured using an inversion recovery pulse sequence in which a $180^\circ$ hard pulse is first applied, followed by a variable delay time before a $90^\circ$ pulse and the signal is acquired. $T_1$ is then calculated from the plot of signal against delay time. Both poppy and millet seeds have rather short $T_1$, i.e., 200 ms, and only 32, i.e., RARE factor, echoes are collected per excitation. The field of view (FOV) is 6 cm in the $x$, 12 cm in $y$ and 3 or 6 cm (depending on the system being 50\% filled or higher) in the $z$-direction. The resolution is thus 0.94 $\mbox{mm}\ \mbox{pixel}^{-1}$, giving SNR = 10 after four averages. The echo time is 5.1 ms, with a 300 ms repetition time  and the whole acquisition time of just under 4 min for a $128 \times 64 \times 64$ image. It is also noted that the magnetisation is returned to the origin of k-space after each RARE readout sequence. This means that all k-space lines are acquired in the same direction and the odd-even echo problem is thus reduced \cite{mantle03}.

Radial and axial segregation are also captured in real time by using 2D FLASH. The pulse used is a 256 ms Gaussian pulse with an excitation pulse angle of $10^\circ$. The echo time TE = 1.3 ms and repetition time TR = 4.7 ms. As for radial segregation, an 82\% full cylinder, rotated at 10 rpm is chosen. Radial segregation occurs quite rapidly so only one average is used to gain temporal resolution. In order to gain enough SNR, a 5 cm thick slice in the $xz$ plane is imaged due to negligible axial segregation within the first revolutions, with the FOV of 6 cm $\times$ 6 cm so as to avoid fold over. The in-plane resolution is chosen to be $64 \times 64$, resulting in temporal resolution of 300 ms. As for axial segregation, the fuller the cylinder, the slower segregation. Therefore, only a 50\% full cylinder is imaged so as to keep segregation within practically short times, i.e., $\sim$ 13.5 min instead of many hours. A 2 mm thick slice in the $yz$ plane, running along the rotation axis and cutting through the center of the sample, with the FOV of 12 cm $\times$ 3 cm, resolution of 128 $\times$ 32 is imaged, repetition time of 150 ms. Due to the thin slice used and the inherent low SNR of FLASH, four averages are used to collect enough signal in this case. FLASH uses a single gradient echo to acquire data, which reduces the repetition time to a few milliseconds per k-space line. The main disadvantage of FLASH is the inherent low SNR since the signal is only proportional to $\sin\theta$. However, the longitudinal magnetisation is then proportional to $\cos\theta$ and is mostly compensated by spin lattice ($T_1$) relaxation, especially in short $T_1$ systems, thus becomes saturated quickly and an infinite number of images could, in principle, be acquired without losing SNR \cite{mantle03}.

\begin{figure} [h]
\begin{center}
\includegraphics[width=0.8\columnwidth]{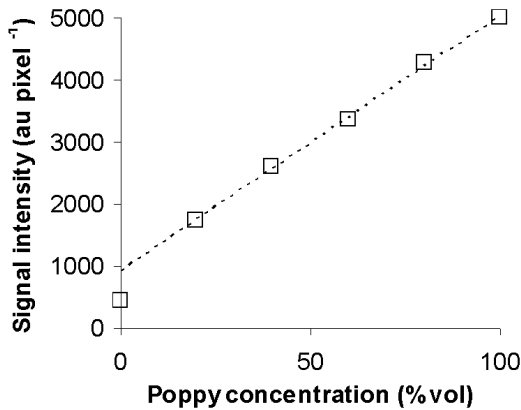}
\caption{\label{fig4}Calibration of the signal intensity as a function of the poppy concentration by volume for the 3D RARE sequence proposed.}
\end{center}
\end{figure}

In all MRI images, the signal intensity in each pixel is obtained in terms of an arbitrary unit (au pixel$^{-1}$) coming from both the poppy and millet seeds and is thus proportional to the amount of seeds present. A calibration of the signal intensity as a function of the poppy concentration by volume is shown in Fig.\ \ref{fig4}. It can be seen that the signal intensity is linearly proportional to the poppy concentration except when this falls below 20\% by volume. This could be attributed to the low signal intensity of the millet seeds (i.e., 500 au pixel$^{-1}$) compared to that of noise (i.e., 195 au pixel$^{-1}$). Nevertheless, the local mixture composition in each pixel can be accurately quantified directly from the signal intensity.

All MRI data are processed using an in-house software package. The raw data (after Fourier transform) are gated at 195 au pixel$^{-1}$ to suppress the background noise. Segregation is found to be rather symmetrical about the rotation axis, and radial segregation in particular is also invariant along the cylinder length. Therefore, projected images can be used to study radial and axial segregation for representative data sets. In addition, signals towards the ends of the receiving coil (in the $y$-direction) start to drop due to the magnetic field inhomogeneity. Therefore probe correction is required for all data sets by using correction factors worked out from a phantom filled with seeds. No correction is necessary in the $x$ or $z$ direction.

\section{\label{sec:level1-3}Results and Discussions}
\subsection{\label{sec:level2-1}Radial segregation}
In this section, we report core formation during radial segregation in cylinders of 75\% and 82\% filling. Based on the core size, the avalanche layer thickness can be inferred, which is then compared with measured values stated in the literature. In addition, the role of core diffusion is confirmed by examining the poppy concentration in the core during radial segregation. Also, the mechanism of initial radial segregation is addressed by looking at real-time 2D FLASH movies. Such full systems have been chosen partly due to the slower rate of radial segregation and partly because core formation occurs.

\begin{figure*}
\includegraphics[width=0.9\columnwidth]{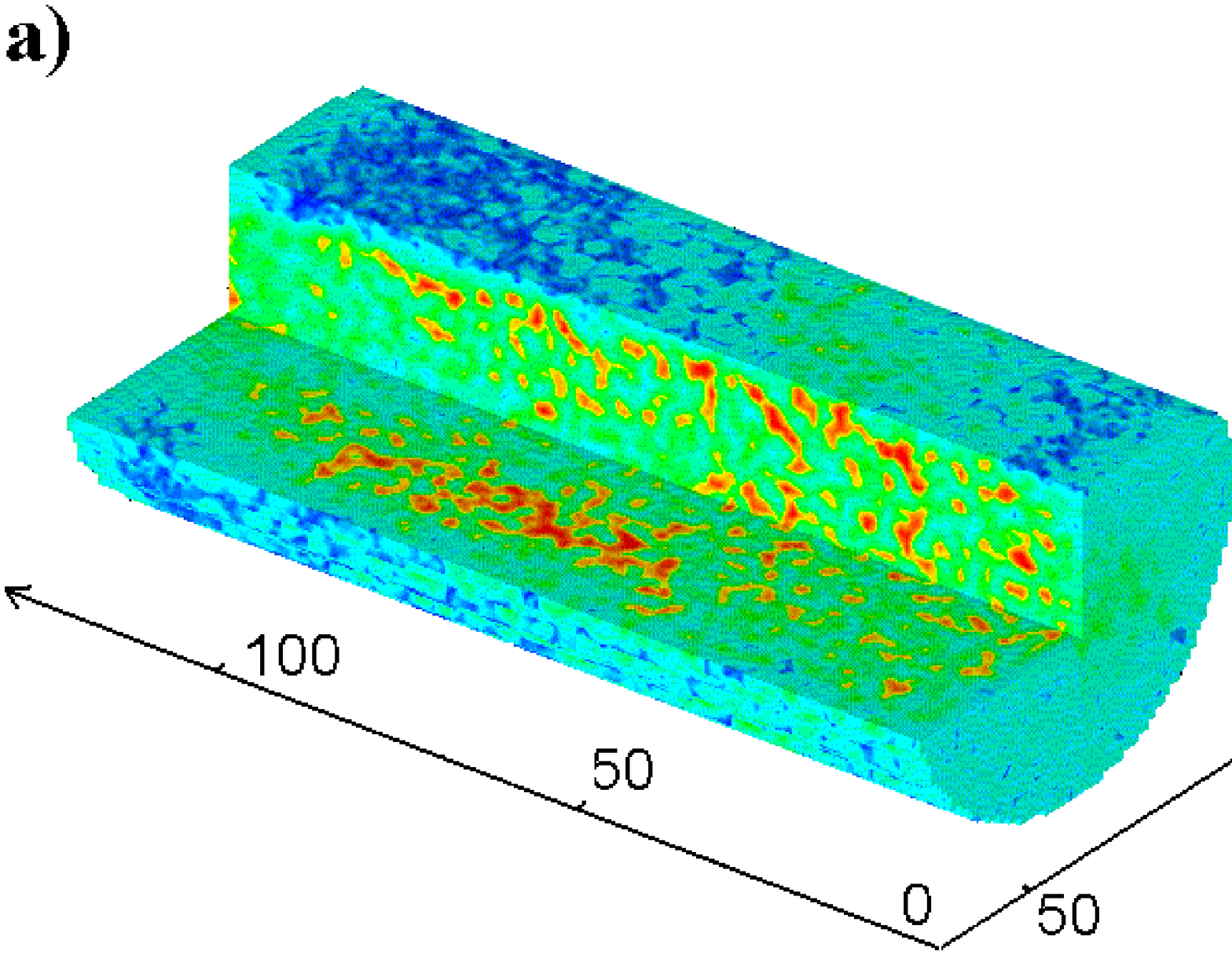}
\includegraphics[width=1.08\columnwidth]{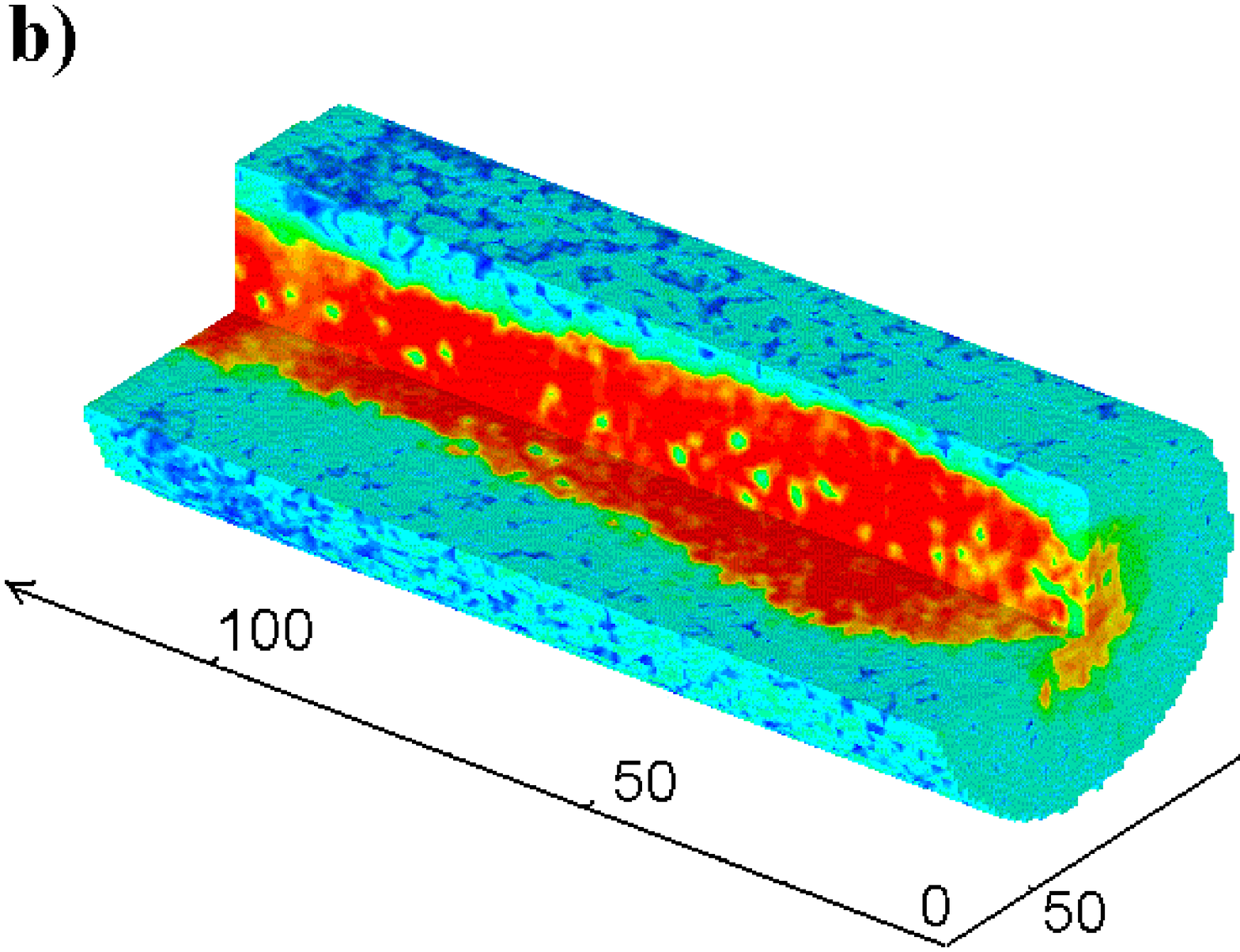}
\caption{\label{fig5} 3D structure of radial segregation in a 75\% full cylinder after (a) 0 s and (b) 960 s of rotations at 35 rpm. Color codes: red (poppy), green (mixture of poppy and millet), cyan (millet), and blue (void). FOV $x \times y \times z$: 6 cm $\times$ 12 cm $\times$ 6 cm. The coordinates are in mm.}
\end{figure*}

Figure \ref{fig5} shows the 3D structure of a 75\% full and 15 cm long cylinder, rotated at 35 rpm, prior to and after 960 s (560 revolutions) of rotation. The concentration of the poppy seeds is linearly proportional to the signal intensity, hence can be easily calculated as shown in section \ref{sec:level1-2}. The color bar shows the corresponding poppy concentration for each color, however for easy identification, red indicates poppy-rich region, green is an equal mixture of poppy and millet, cyan is millet, and blue is void. It can be seen that seeds are initially well-mixed with mostly green color in the cut-through view. Radial segregation only comes to completion after 960 s when most of the poppy seeds concentrate in a cylindrical core along the entire length of the rotation axis, whereas the rest of the drum is millet, i.e., cyan color.

\begin{figure*}
\includegraphics[width=0.62\columnwidth]{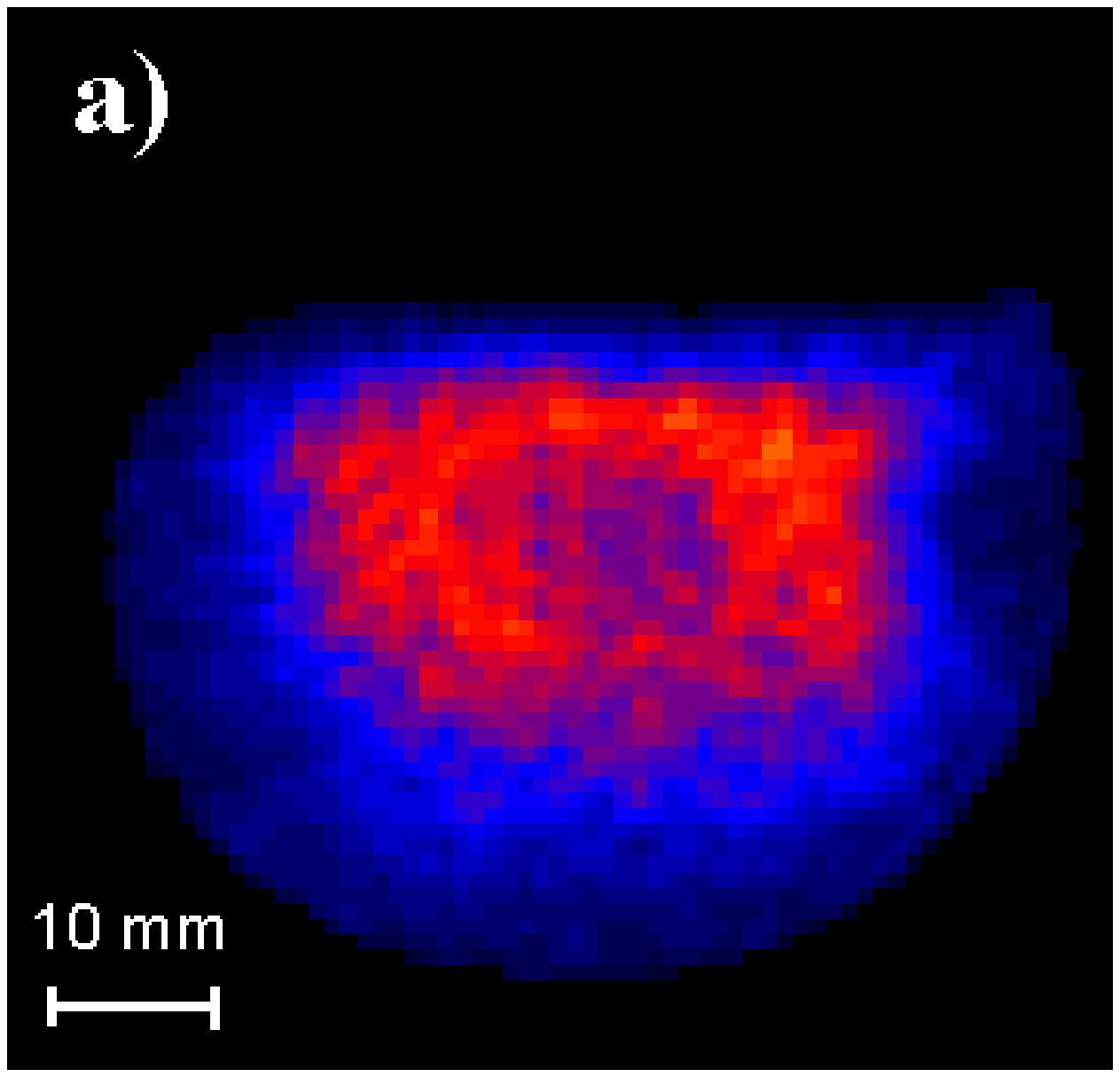}
\includegraphics[width=0.61\columnwidth]{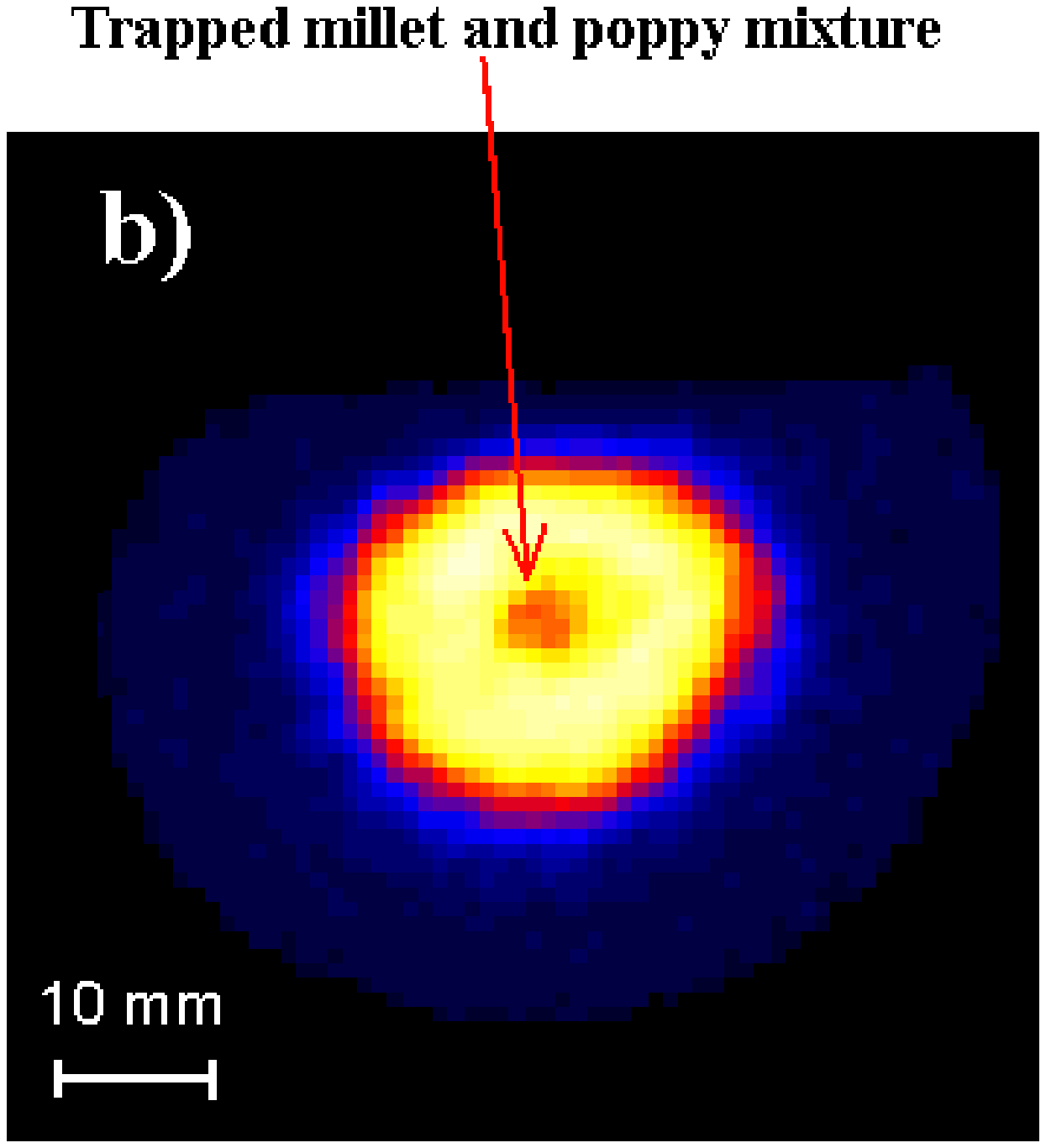}
\includegraphics[width=0.71\columnwidth]{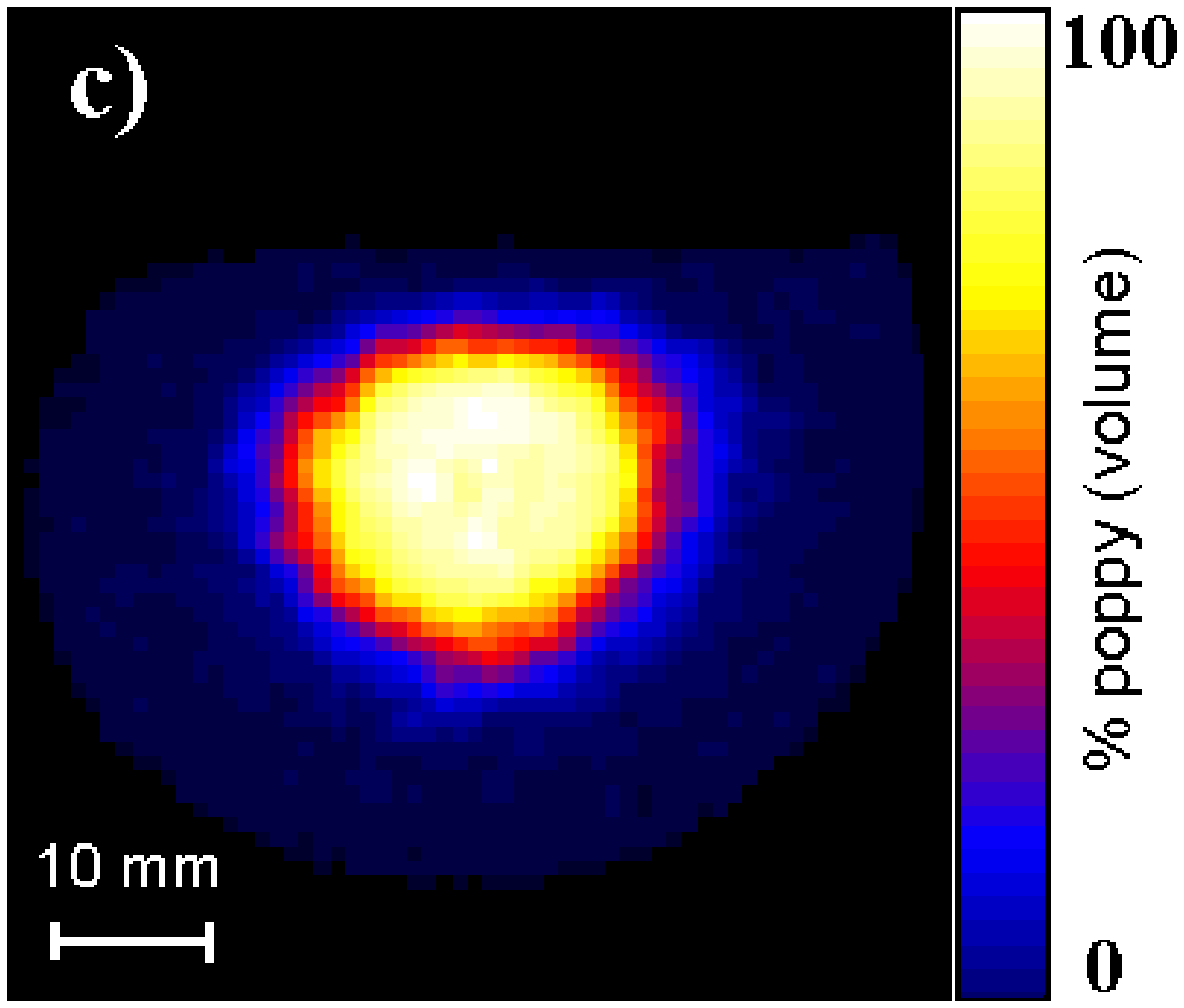}
\caption{\label{fig6} 2D $xz$ projection, produced from the 3D structures in Fig.\ \ref{fig5} showing radial segregation in a 75\% full cylinder after (a) 2 s, (b) 180 s and (c) 960 s of rotations. Color codes: brighter colors indicate higher poppy concentration and darker colors indicate higher millet concentration. FOV $x \times y$: 6 cm $\times$ 6 cm.}
\end{figure*}

\begin{figure*}
\includegraphics[width=\columnwidth]{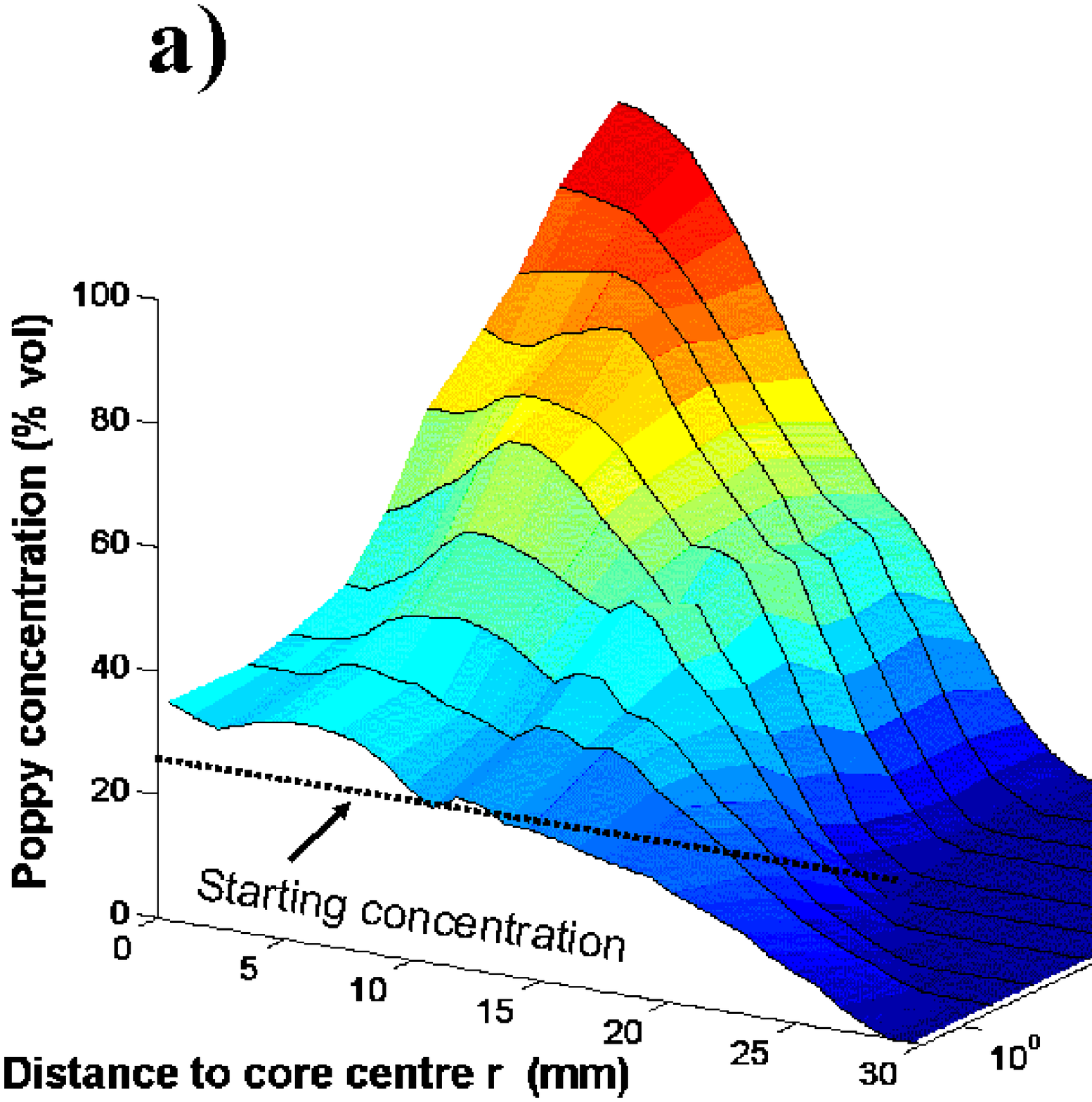}
\includegraphics[width=\columnwidth]{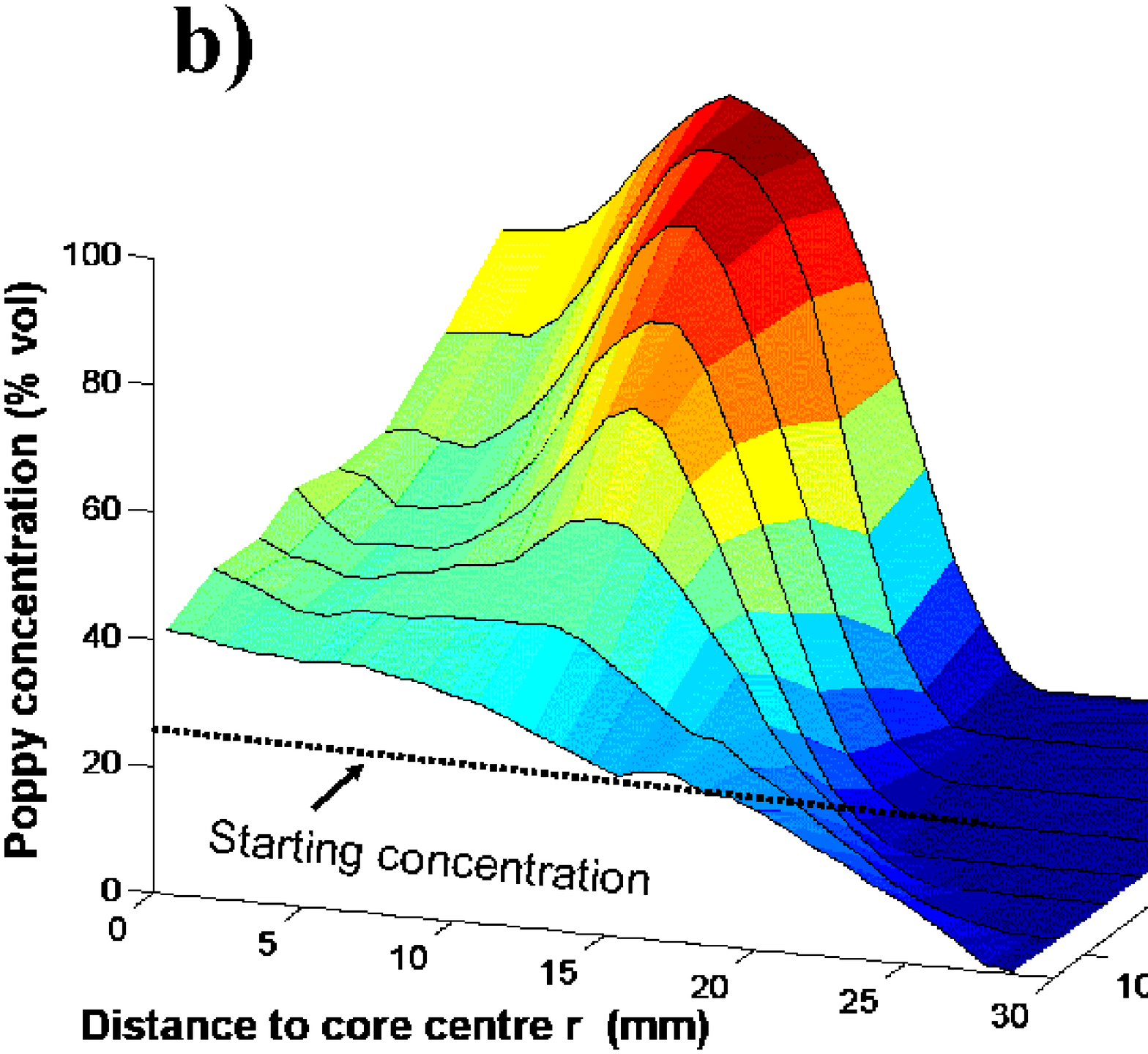}
\caption{\label{fig7} 1D concentration profile across the cylinder radius in (a) 75 and (b) 82\% full cylinder during radial segregation at 35 rpm. Zero distance from core center indicates the rotation axis whereas maximum distance means the cylinder wall.}
\end{figure*}

A series of these stationary 3D images are acquired at many intermediate rotations showing the progress of radial segregation. Each 3D image is then collapsed into a 2D projection in the $xz$ plane due to negligible axial segregation such as those shown in Fig.\ \ref{fig6}. Since radial segregation concerns mainly the traveling of seeds radially and the 2D projection is rather symmetrical about its center point, it is possible to comprehensively represent the status of radial segregation by a 1D profile across the cylinder radius. This is done by averaging the poppy concentration at each distance from the core center in the 2D projection to get a 1D profile as shown in Fig.\ \ref{fig7}(a). A similar procedure is also done with an 82\% full, 12 cm long cylinder, and results are shown in Fig.\ \ref{fig7}(b). A zero distance to core center indicates the center point of rotation, i.e., on the rotation axis; whereas a maximum distance, i.e., 30 mm away, means the cylinder wall.

At both filling fractions, it is found that subject to rotation, poppy seeds immediately move inward, forming a core so that only millet seeds are visible from the outside. This can easily be seen in Fig.\ \ref{fig6}(a) after 2 s of rotation, or in Fig.\ \ref{fig7} where there is, as soon as rotation starts, an abrupt drop of the signal intensity near the cylinder wall, i.e., r = 30 mm. Further rotations lead to the formation of an annulus of poppy seeds, which is visible at 180 s, trapping some of the millet seeds at the core center, which corresponds to approximately 50\% poppy by volume, in Fig.\ \ref{fig6}(b) and a trough in the 1D profile in Fig.\ \ref{fig7} at the core center. The maxima in all 1D profiles correspond to the most poppy-rich region inside the cylinder. 

From this point in time, the behaviors of the 75\% and 82\% full cylinders diverge. In the case of the 75\% full cylinder, the maximum of the 1D profile is only temporary and disappears as radial segregation progresses to its end. This can be clearly seen in Fig.\ \ref{fig6}(c) in which the poppy and millet mixture in the core in Fig.\ \ref{fig6}(b) has disappeared before the 960 s has elapsed. Equivalently, Fig.\ \ref{fig7}(a) shows that this poppy-rich region initially appears at 8 mm away from the core center after a few seconds of rotations but gradually migrates towards the core center until they meet after 250 s of rotations. Subsequently, poppy seeds at the core center keep accumulating until its concentration reaches 100\%. In the case of the 82\% full system, however, the maximum of the 1D profile is clearly permanent as it persists through to the completion of radial segregation, which is also the onset of axial segregation after 2100 s of rotation as shown in Fig.\ \ref{fig7}(b).

This formation and disappearance of the core center during radial segregation are not only interesting dynamic behaviors in themselves but also tell us some information about the avalanche layer thickness. 

Firstly, for a 75\% full cylinder, the disappearance of the core center means that the avalanche layer extends well into the center of rotation. From geometrical considerations, this distance (i.e., from the top of the free surface to the center of rotation) is 12 mm, which means that the avalanche layer in a 75\% full cylinder, rotated at 35 rpm is at least 12 mm. In the literature to date, there is a range of reported values for the free surface layer thickness. One author reported a thickness of 6 mm for a pile of mustard seeds in an avalanche \cite{jaeger92}, in which case the core center in our system is buried well under the free surface. However, it has also been argued that the free surface layer may be thicker than previously thought where depths of 13 mm at 22 rpm \cite{seymour00} and 18 mm at 30 rpm \cite{yamane98}, both in a cylinder of 70 mm in diameter, have been measured by MRI.

Secondly, for an 82\% full cylinder, the core is clearly permanent during radial segregation as mentioned above. The location of the final, stable maximum can be regarded as the boundary between the diffusion-dominant core and the convection-dominant free surface region --- 8.5 mm from the core center. This corresponds to a distance of 7.5 mm from the free surface to the boundary, which can be considered as the avalanche layer thickness. However, it has just been shown in the 75\% full system that the distance from the core center to the free surface is 12 mm. This could be attributed to two reasons. First, the avalanche layer thickness could be dependent on the filling fraction. In the case of an 82\% full, the free surface area is much smaller than that of a 72\%, so that the distance available for acceleration of seeds from the top to the middle of the avalanche layer is smaller, thus the avalanche layer is dimmed thinner. Second, the avalanche layer could have significant effect on the core underneath so that in a 75\% cylinder, the millet seeds initially trapped in the core center have diffused out, leaving a more or less pure poppy core. 

As for the role of core diffusion, we examine the poppy concentration at the core center during radial segregation. In order to separate the effect of the avalanche layer, we only look at the core center in the 82\% full cylinder, which has been suggested above to be buried well under the avalanche layer, i.e. 8.5 mm. As shown in Fig.\ \ref{fig7}(b), this concentration increases in a monotonous fashion from 25\% by volume, i.e., homogeneously mixed state, to 67\%. This indicates an outward flux of the millet seeds, which is most likely to be brought about by core diffusion as no convection is observed in this area. This confirms the importance of core diffusion, along side with the free surface, by which segregation takes place.

\begin{figure*}
\includegraphics[width=0.66\columnwidth]{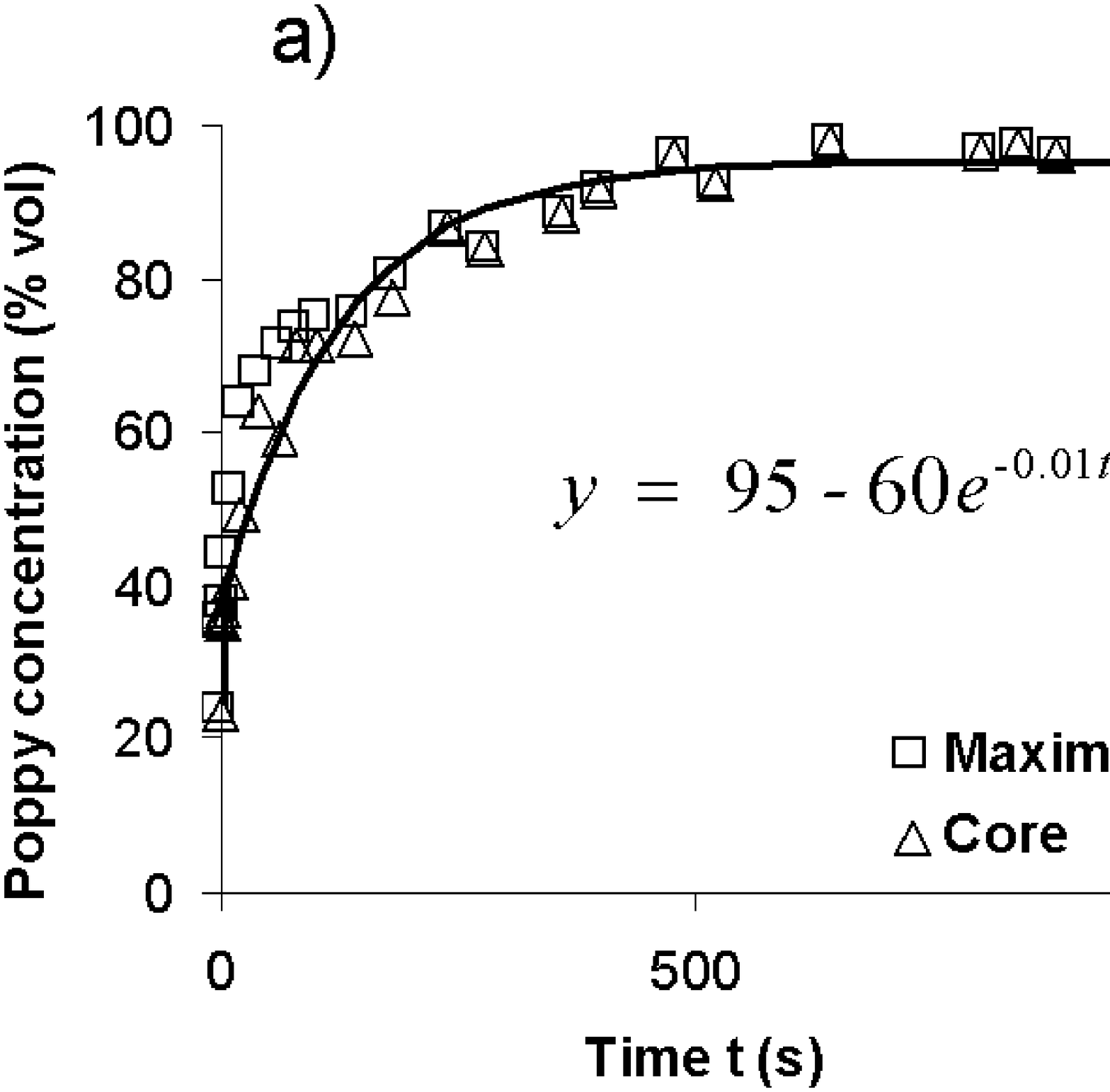}
\includegraphics[width=0.66\columnwidth]{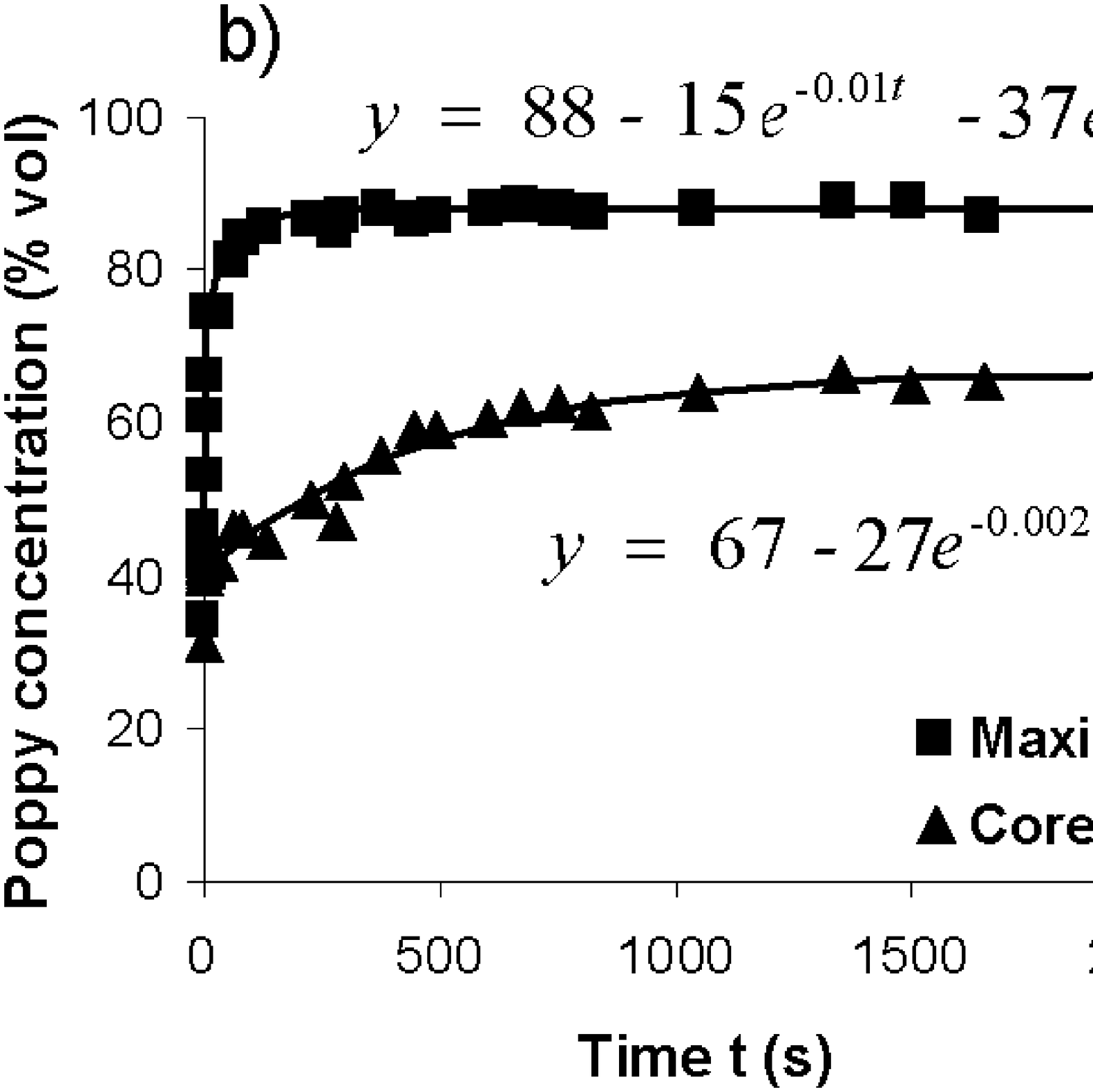}
\includegraphics[width=0.66\columnwidth]{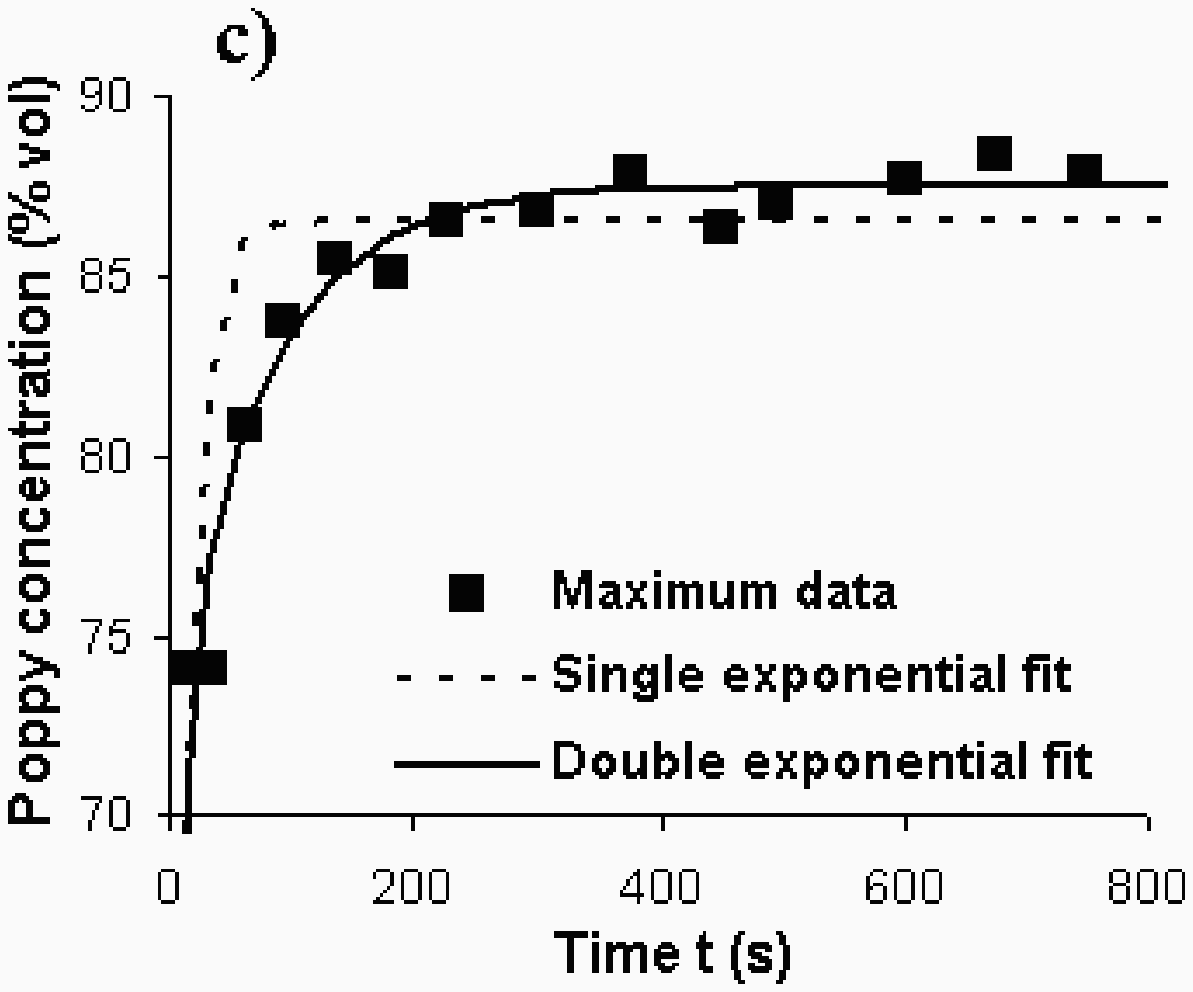}
\caption{\label{fig8} Poppy concentration at the core center and the maximum of the 1D concentration profile across the cylinder radius in Fig.\ \ref{fig7} during radial segregation in (a) 75\% and (b) 82\% full cylinder. (c) Comparison of the single and double exponential fit to the data for the maximum concentration in the 82\% full cylinder.}
\end{figure*}

To assess the relative significance of core diffusion and the free surface, we look into the accumulation rate of poppy seeds at the core center and the most poppy-rich region, i.e., at the maximum of the 1D profile, as plotted in Fig.\ \ref{fig8}. As can be seen in Fig.\ \ref{fig8}(a) for the 75\% full cylinder, the maximum and core concentrations are essentially the same except during the first 200 s when there still exists a millet and poppy mixture core. The poppy concentration in the core increases at an exponentially decaying rate as shown in the best fit equation. The rate at which the poppy seeds accumulate inside the core is represented by the exponential term of 0.01 s$^{-1}$. The 82\% full system, on the other hand, has distinctive signal intensities at the core center and the most poppy-rich region, also defined as the boundary between the core and the avalanche layer, as shown in Fig.\ \ref{fig8}(b), confirming the existence of the millet and poppy mixture in the core. It clearly proves the diffusion of millet seeds, in other words the accumulation of poppy seeds, inside the core at a rate represented by the exponential term of 0.002 s$^{-1}$. In comparison with that of the 75\% full system, i.e., 0.01 s$^{-1}$, it, however, occurs much more slowly due to the fact that the diffusion-dominant core is thicker for the 82\% full system from geometrical considerations. 

The most interesting feature to note is that the accumulative concentration at the poppy annulus follows a double exponential equation. Fitting the data to this double exponential equation yields a perfect fit whereas that to a single exponential equation is not satisfactory as shown in Fig.\ \ref{fig8}(c). The fitted single exponential line shoots up very fast and levels out abruptly at t = 70 s while the experimental data levels out much more gradually. In the double exponential equation, the first term has the same exponential power of 0.01 s$^{-1}$ as the 75\% full system, but the second exponential power is higher, i.e., 0.14 s$^{-1}$, indicating a higher rate of poppy seed accumulation. This suggests two processes going simultaneously at this boundary: diffusion inside the annulus and convection outside it from the avalanche layer. This is another confirmation that radial segregation is brought about by both convection in the free surface and diffusion inside the core. When the system is less than 75\% full, all seeds inside the cylinder reach the avalanche layer in every rotation of the cylinder and thus models and theories based on the dynamics of the avalanche layer are appropriate. However, when the cylinder is full like those presented in this work, diffusion plays a significant role in the physics inside the core along the entire rotation axis.

\begin{figure*}
\includegraphics[width=0.486\columnwidth]{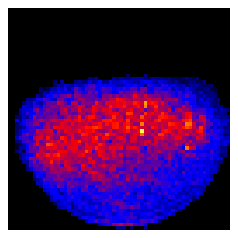}
\includegraphics[width=0.486\columnwidth]{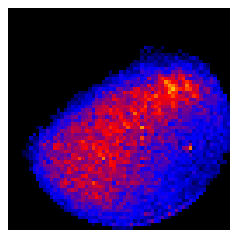}
\includegraphics[width=0.486\columnwidth]{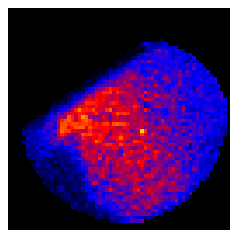}
\includegraphics[width=0.486\columnwidth]{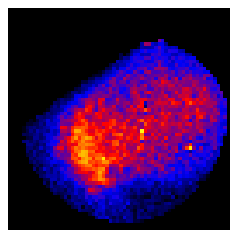}
\includegraphics[width=0.486\columnwidth]{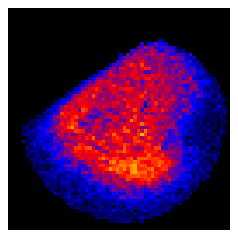}
\includegraphics[width=0.486\columnwidth]{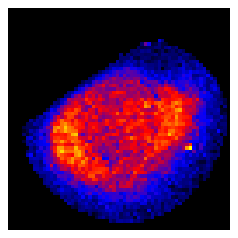}
\includegraphics[width=0.486\columnwidth]{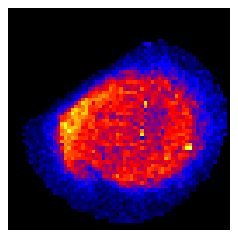}
\includegraphics[width=0.486\columnwidth]{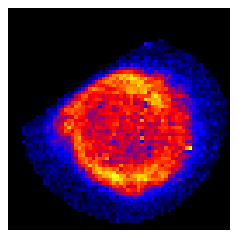}
\caption{\label{fig9} 2D FLASH pictures capturing radial segregation in an 82\% full cylinder, rotated at 10 rpm during the first 12 s of rotation. The timings of these snap shots from left to right, top to bottom are: 0.9, 1.8, 3.0, 4.2, 5.4, 6.6, 9.0 and 13.8 s. Color codes: blue: millet, red: mixture, yellow-white: poppy.}
\end{figure*}

It should be again noted that all 3D structure images presented so far are acquired when the system is at rest. Radial segregation in an 82\% full cylinder, rotated at 10 rpm is also captured in real time by 2D FLASH MRI and the movie can be viewed at the provided link \cite{movie1}. Some pictures extracted from this movie, showing initial radial segregation within the first 13.8 s of rotation, can be seen in Fig.\ \ref{fig9}. It is evident that initial formation of the poppy core during radial segregation within a few rotations is due to the stopping and percolation of poppy seeds, as suggested by Ref.\ \cite{cantelaube95}. \\[2mm]

Radial segregation: 82\% full cylinder at 10 rpm: \cite{movie1}

\subsection{\label{sec:level2-2}Axial band formation}
We present 3D MRI data for sections of a 50\% full, 50 cm long drum, rotated at 35 rpm. The progress of the band formation is captured, thus its rate can be calculated. Axial band formation is a familiar topic but we demonstrate in this section that full 3D structure of the cylinder can be captured by MRI, thus is useful for modeling work. We will also address the role of core diffusion. For this purpose, we will present MRI data for an 82\% full system. 

\begin{figure*}
\includegraphics[width=0.9\columnwidth]{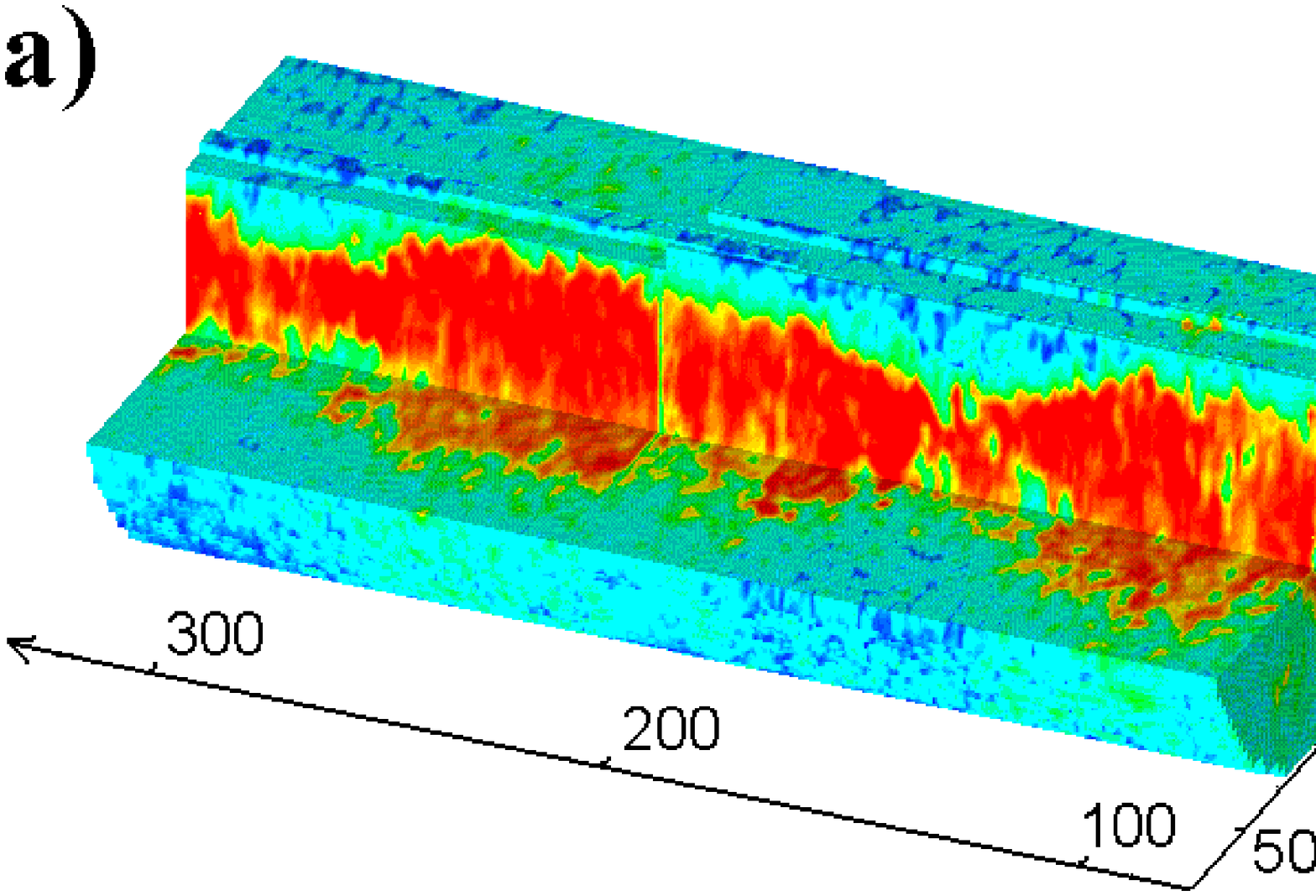}
\includegraphics[width=1.08\columnwidth]{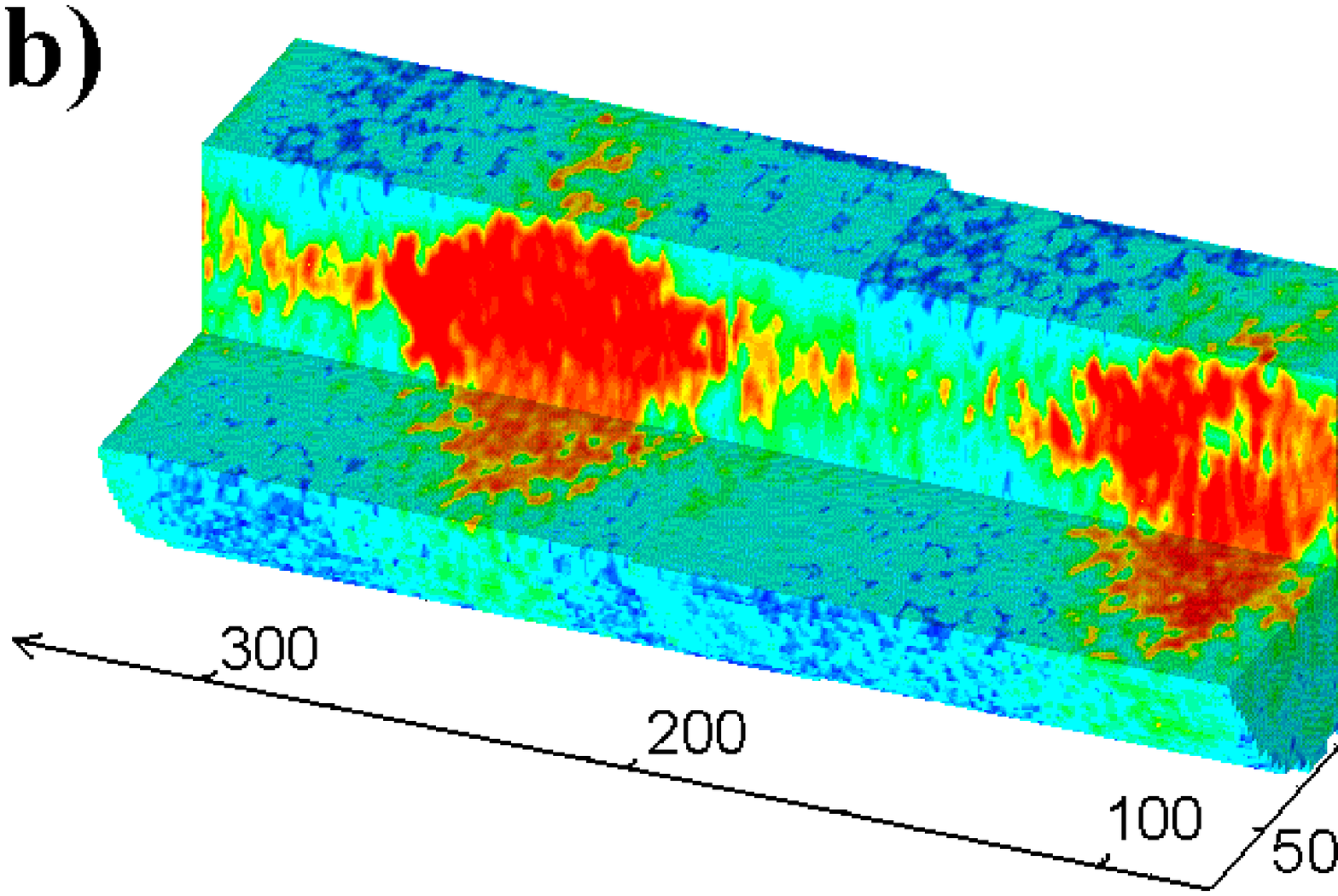}
\caption{\label{fig10} 3D structures showing the formation of two axial bands in a section near one end of a 50\% full cylinder after (a) 270 s and (b) 899 s of rotations at 35 rpm.}
\end{figure*}

Figure \ref{fig10} shows the progress of axial segregation in a 22 cm section near one end of the rotating drum, which is made by putting together two FOV 12 cm images end-to-end (overlapping 2 cm). There is a small band right at the end cap (toward the right of the images at $y$ = 0), which is not shown but we capture the formation of two real bands adjacent to it. At 270 s, the bands start to form with poppy seeds breaking off at two locations along the rotation axis. At 899 s, the formation of two axial bands is more or less complete and it is evident that there are poppy seeds along the entire axis, ranging from 40\% to 60\% by volume as indicated by the red and green colors, connecting the two poppy bands. 

\begin{figure}
\includegraphics[width=\columnwidth]{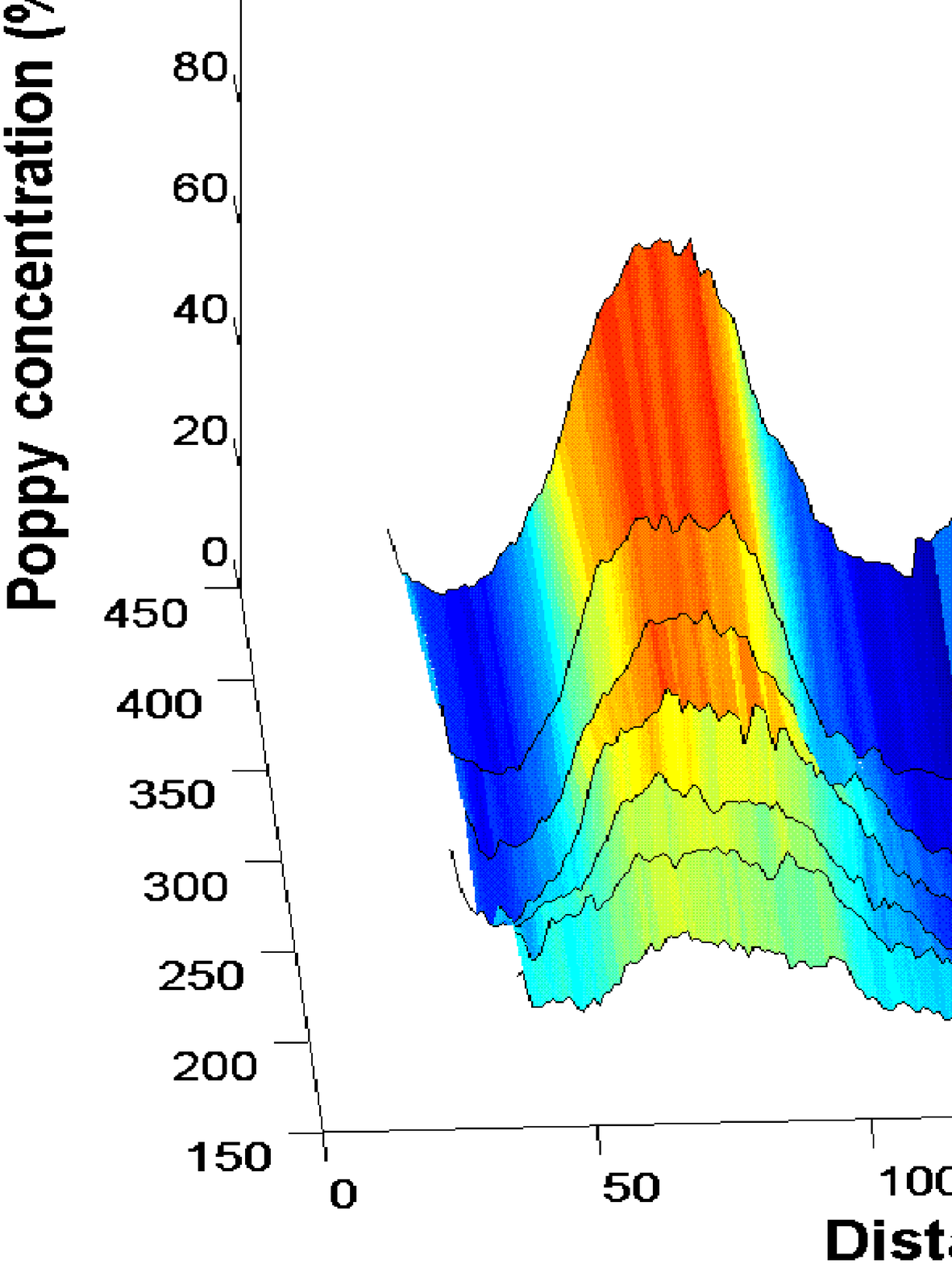}
\caption{\label{fig11} Axial band formation in a 22 cm section of a 50\% full and 50 cm long cylinder, rotated at 35 rpm.}
\end{figure}

Similar to the procedure employed for analyzing radial segregation, a series of 3D images like those in Fig.\ \ref{fig10} are acquired and collapsed into 1D profile showing the concentration of poppy seeds along the rotation axis by averaging the signal intensity along the $x$ and $z$ axes due to their symmetry. Results are shown in Fig.\ \ref{fig11}. During the course of band formation, the band position does not change, but only shortens in width and increases in height, i.e., poppy concentration. Axial band formation therefore involves essentially the diffusion of millet seeds from the nascent poppy band to the adjacent nascent millet band, which is in agreement with the mechanism by the dynamic angle of repose as proposed by Kakalios \cite{kakalios05}. The rate at which these two bands are formed can be determined by fitting an equation to the bands' maximum poppy concentration. An increase of poppy accumulation at an exponentially decaying rate similar to Fig.\ \ref{fig8} is observed. 

\begin{figure}
\includegraphics[width=\columnwidth]{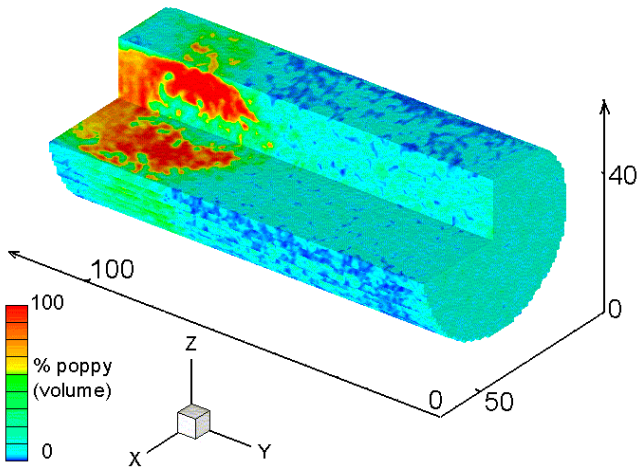}
\caption{\label{fig12} 3D structure of the final state of axial segregation in an 82\% full and 11 cm long cylinder, rotated at 35 rpm.}
\end{figure}

It is of interest to examine if the dynamics of the avalanche layer is the only mechanism of axial segregation, as claimed in many studies. For example, an attempt to verify this mechanism was made by conducting a simple experiment, in which a co-axial cylinder was inserted into the segregating cylinder and axial segregation was no longer observed \cite{zik94}. The author then concluded that free surface was the only mechanism for segregation since the experiment eliminated free motion along the surface. However, this setup also removed the opportunity for core formation; hence a conclusion cannot be drawn about the sole mechanism of segregation. Instead, we conduct a different experiment, in which a rather full cylinder is employed. If the free surface is the only mechanism, it is anticipated that axial segregation should not occur in such a full system since there is only a small amount of free surface and more importantly, there exists a core which is buried under the free surface as presented previously when discussing radial segregation. The cylinder to be studied is the same as that for the radial segregation, which is 12 cm long and filled to 82\% by volume. The final state of this system after 55,000 revolutions (26 hours) is shown in Fig.\ \ref{fig12}. The system still segregates like a 50\% system so that there is one poppy band next to the left end cap, and the rest is filled with millet seeds. As we saw earlier, there exists a core along the entire length of the rotation axis with the diameter of at least 7.5 mm buried under the avalanche layer. However, this core only remains inside the poppy band and has completely disappeared from inside the millet band. This is a clear indication of core diffusion leading to complete axial segregation, which is in line with the sub-surface axial segregation observed previously \cite{hill97}. This core of millet and poppy mixture is not visible to the eye or by ordinary measurements but can be detected by MRI. As shown in the cut-through view in Fig.\ \ref{fig12} there is a core of $\sim$40\% poppy, located on the rotation axis but inside the poppy band.

\begin{figure}
\includegraphics[width=0.8\columnwidth]{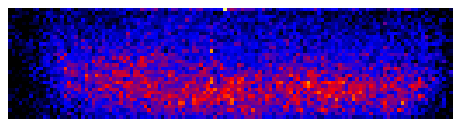}
\includegraphics[width=0.8\columnwidth]{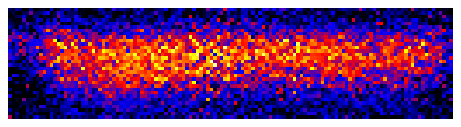}
\includegraphics[width=0.8\columnwidth]{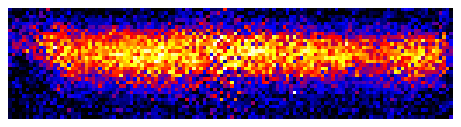}
\includegraphics[width=0.8\columnwidth]{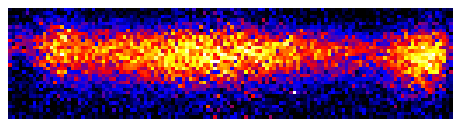}
\includegraphics[width=0.8\columnwidth]{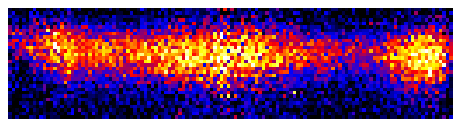}
\includegraphics[width=0.8\columnwidth]{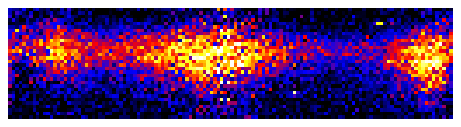}
\caption{\label{fig13} Axial band formation in a 50\% full and 12 cm long cylinder, rotated at 35 rpm during the first 2.5 minutes of rotation. The timings of these snap shots from the top to bottom are: 0.3, 3.6, 18.8, 95.0, 133.7 and 284.4 s (4.7 min). Color codes: blue: millet, red: mixture, yellow-white: poppy.}
\end{figure}

Real time movies showing the band formation can also be acquired by 2D FLASH MRI. A 50\% full and 12 cm long system is chosen so as to keep the segregation time sufficiently short for data acquisition. The movie \cite{movie9} is recorded over 4.7 min of axial segregation and some intermediate pictures are shown in Fig.\ \ref{fig13}. In this movie, one can clearly see the formation of two poppy bands at the two end caps and one band in the middle of the cylinder. This is just to demonstrate the applicability of MRI for studying the dynamics of granular materials and that it can provide useful data for modeling work. \\[2mm]

Axial band formation: 12 cm long and 50\% full cylinder: \cite{movie9}

\subsection{\label{sec:level2-3}Axial band traveling and merging}
Upon ``full'' axial segregation, bands have been observed to travel and merge over a long time scale (many hours) by some authors \cite{newey04,choo98,aranson99,fiedor03}. This phenomenon is observed in our study on both short (within 13.5 min) and long (after 33 hours) time scales in a series of successive merging events. In order to keep the time scale sufficiently short for data acquisition, only 50\% full systems are considered. The long time scale band traveling and merging are first studied in a system of a 50 cm long rotating drum by 3D stationary RARE. Such a long cylinder is chosen so as to permit the formation of many axial bands (8 poppy bands in this case), followed by band traveling and merging events. The real-time 2D FLASH movies of both band traveling and band merging in a 12 cm long cylinder are subsequently presented. This short cylinder length is chosen for 2D FLASH so that both band traveling and merging occur within 13.5 min of rotation and are feasible to produce a movie instead of over 33 hours. 

\begin{figure}
\includegraphics[width=0.48\columnwidth]{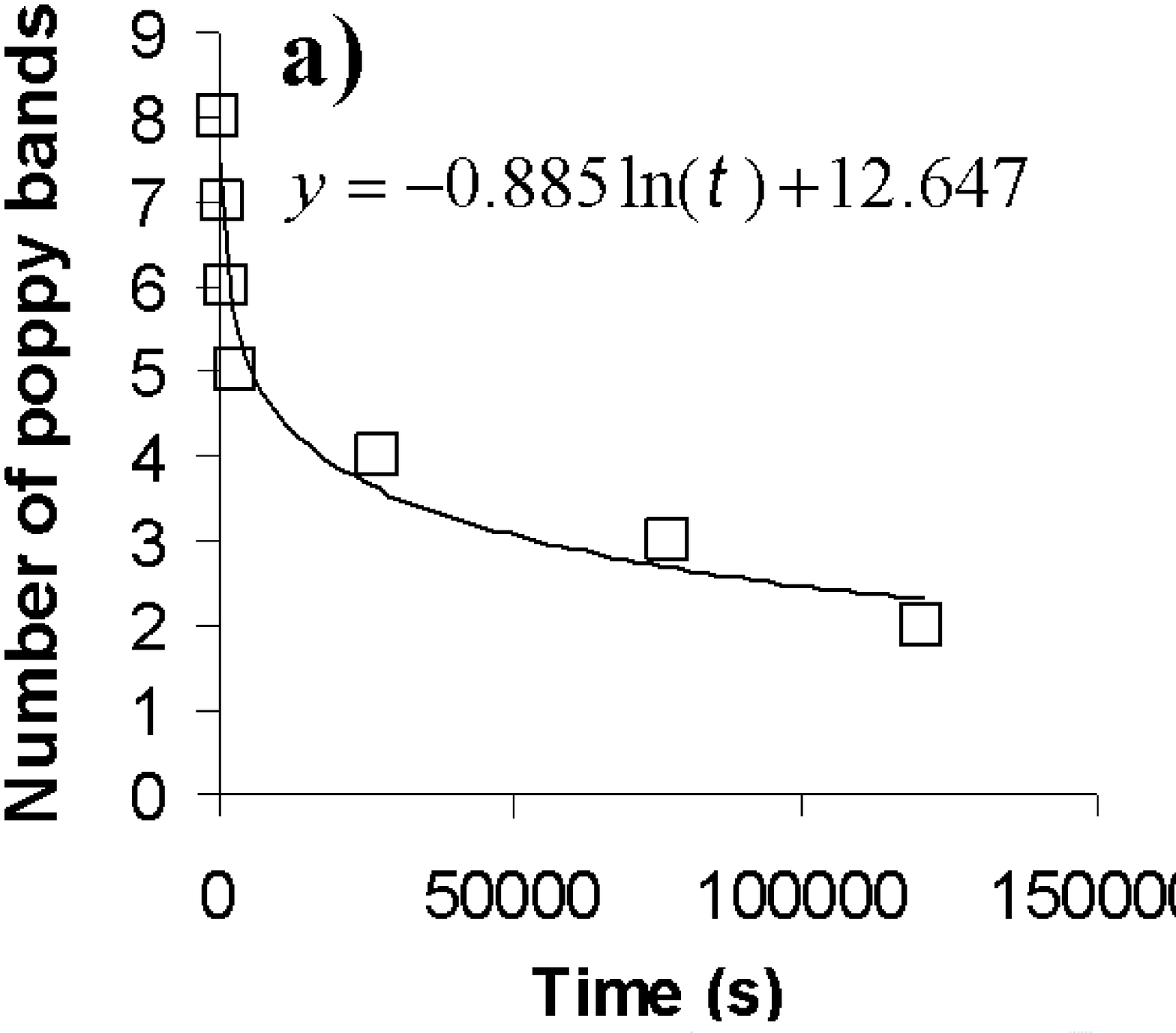}
\includegraphics[width=0.48\columnwidth]{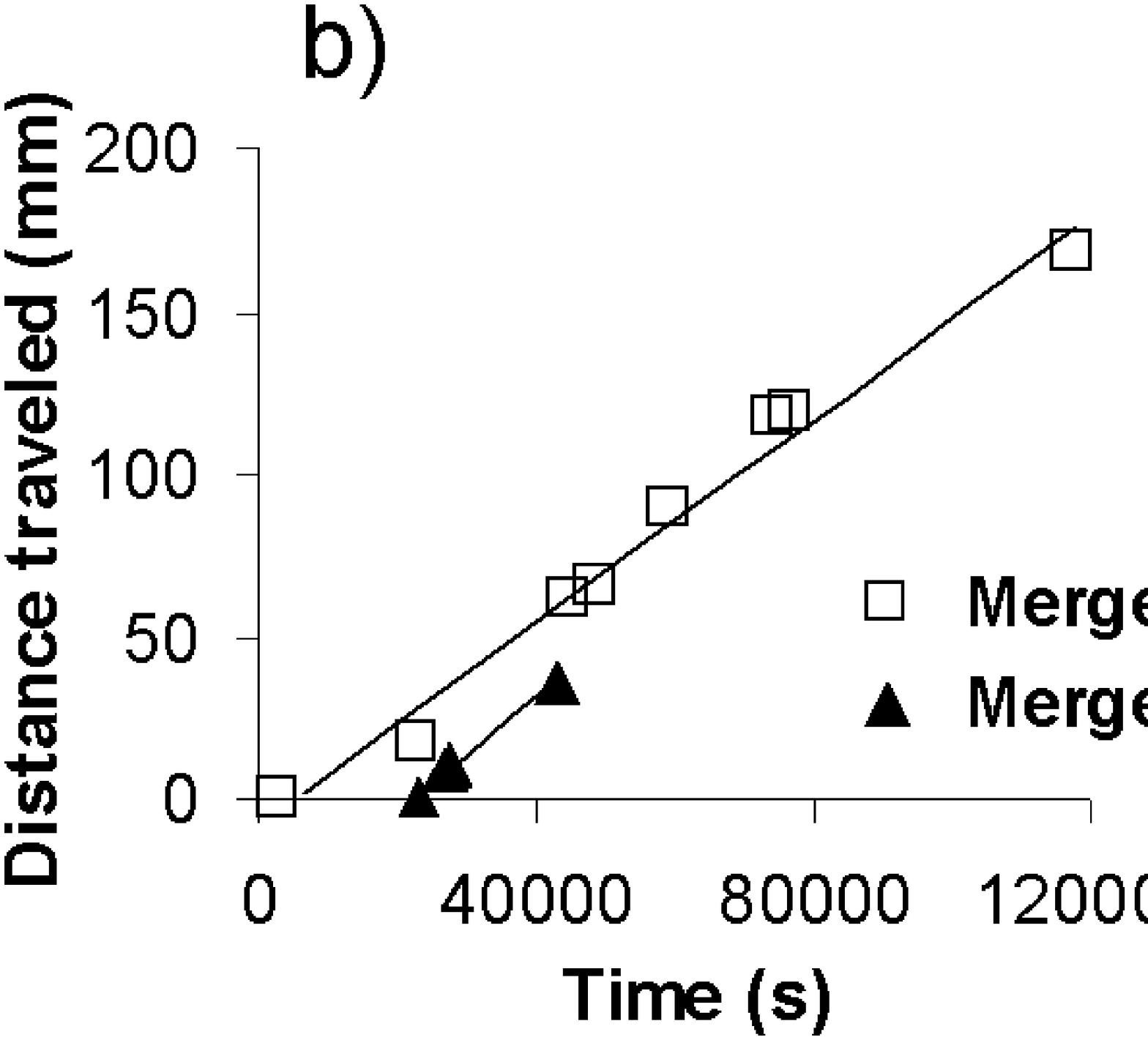}
\caption{\label{fig14} In a 50\% full and 50 cm long cylinder: (a) number of bands over time due to band merging and (b) band traveling prior to two separate merging events.}
\end{figure}

As mentioned previously, there are 8 bands of poppy seeds formed at equal spacings of 7 cm at full axial segregation in a 50\% full and 50 cm long drum. Further rotations over the next 70,000 revolutions or 33 hours result in these bands to merge pair-wise until there are only two bands in the entire cylinder, one is the small original band at one end of the cylinder and the other is a long band located at the other end. The number of poppy bands over time follows a logarithmic decaying rate, as shown in Fig.\ \ref{fig14}(a), which was also observed by Ref.\ \cite{aranson99}.

There are many merging events, which are always preceded by band traveling, starting from 8 poppy bands at full axial segregation, only two of which are chosen as the case study for band traveling. It is observed that prior to each merging event, mainly one band is ``active'' and travels gradually toward the other band, which remains essentially stationary. During traveling, bands maintain their shape (band with side bulges as shown in Fig.\ \ref{fig10}(b)) and size. Figure \ref{fig14}(b) shows the displacement of the traveling bands in the two separate merging events mentioned above over time. The linearity of both curves in Fig.\ \ref{fig14}(b) indicates a reasonably constant, as previously reported \cite{choo98}, but very sluggish traveling speed of 3 $\pm\ 0.2\ \mu\mbox{m}\ \mbox{s}^{-1}$ (taken from the gradient of the two curves). However, once the bands are adequately close --- approximately 3 cm apart --- band merging occurs much more rapidly.

Unlike band traveling, poppy bands remain stationary during merging but change their shape by the movement of millet seeds through the poppy bands. The merging of poppy bands can be seen in Fig.\ \ref{fig15} and summarized as follows. Poppy seeds first travel along the rotation axis from the existing bands, forming a thin line along it, connecting the two bands. This can be seen in Fig.\ \ref{fig15}(a) where the area along the rotation axis between the bands appears green, indicating a poppy and millet mixture consisting of 30-60\% poppy by volume. Millet seeds then keep traveling out of this area through the poppy bands over many stages such as one in Fig.\ \ref{fig15}(b) until the concentration of poppy seeds in the bands and the area between are equal as shown in Fig.\ \ref{fig15}(c). A series of images like those are recorded to confirm the merging mechanism.

These images confirm a similar claim in Ref.\ \cite{hill97} that the merging of bands is not simply their traveling until they meet each other, but involves the removal of millet seeds from between the poppy bands. In that paper, the camera technique was used so only observations on the surface could be made. However, in this paper, full 3D images of particle distribution are available, showing the same process happening in the bulk materials. Also, useful information listed below, such as the merging mechanism and rate, can be extracted: 

\begin{figure*}
\includegraphics[width=\columnwidth]{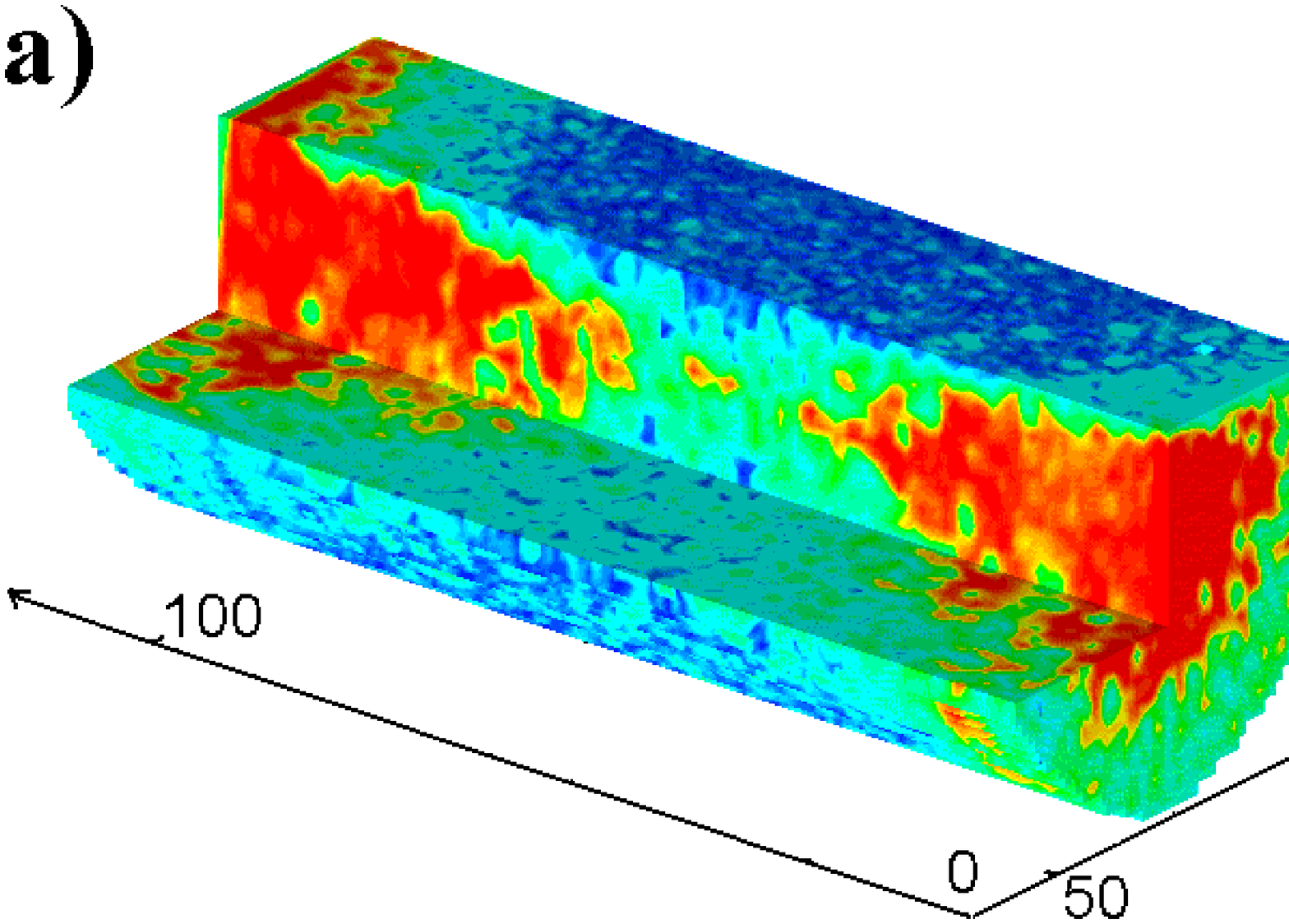}
\includegraphics[width=\columnwidth]{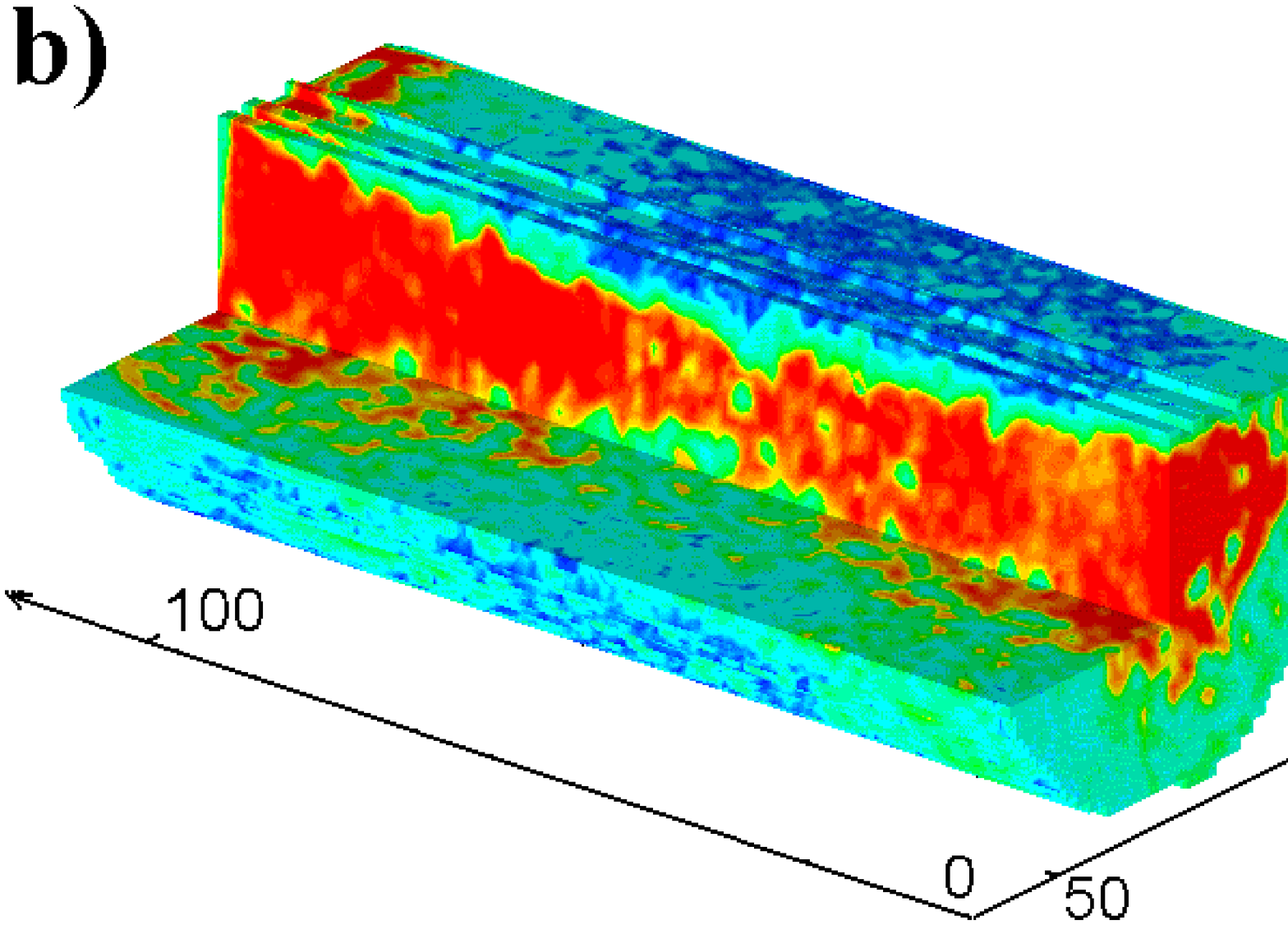}
\includegraphics[width=\columnwidth]{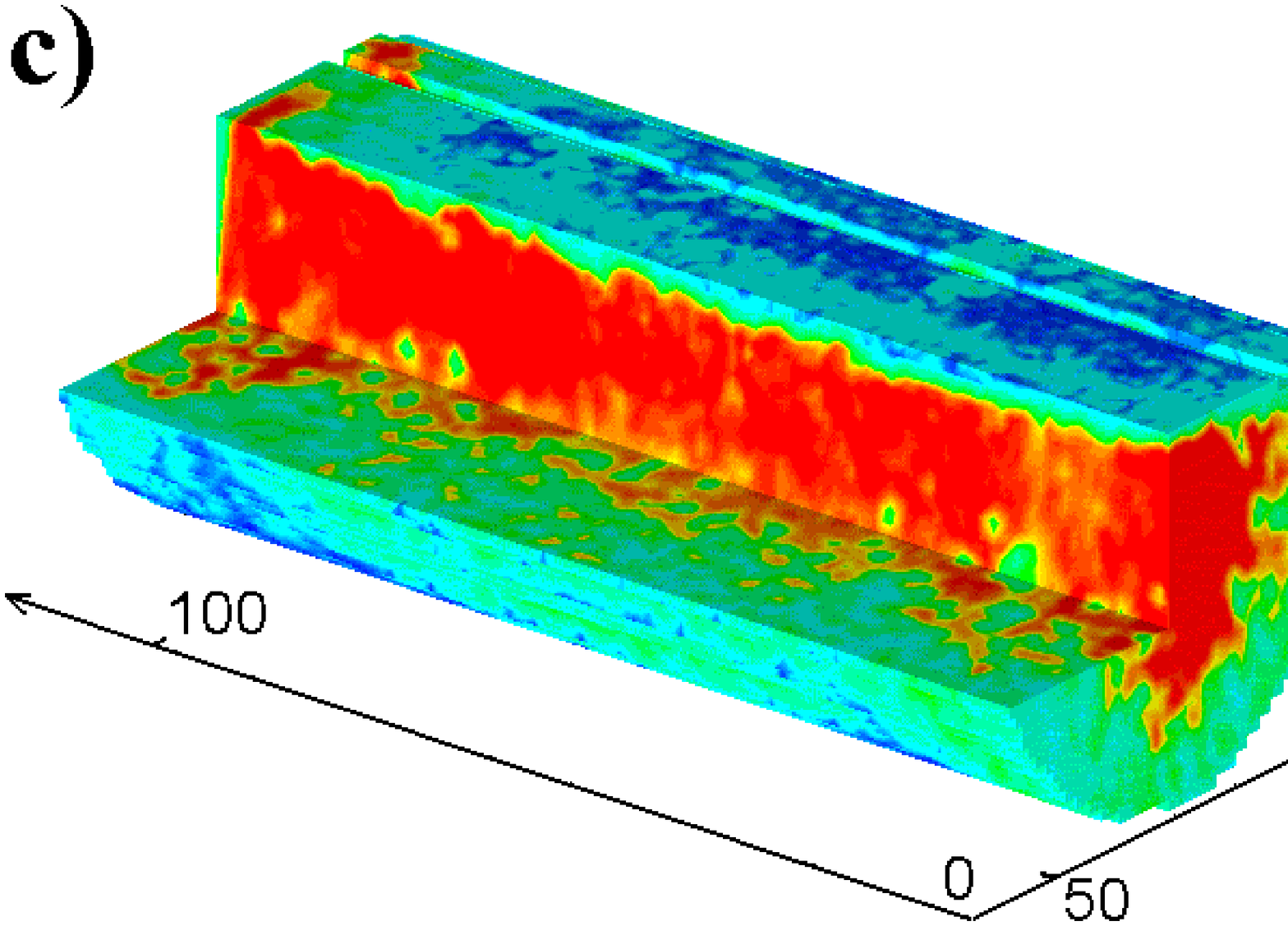}
\hfill \includegraphics[width=0.8\columnwidth]{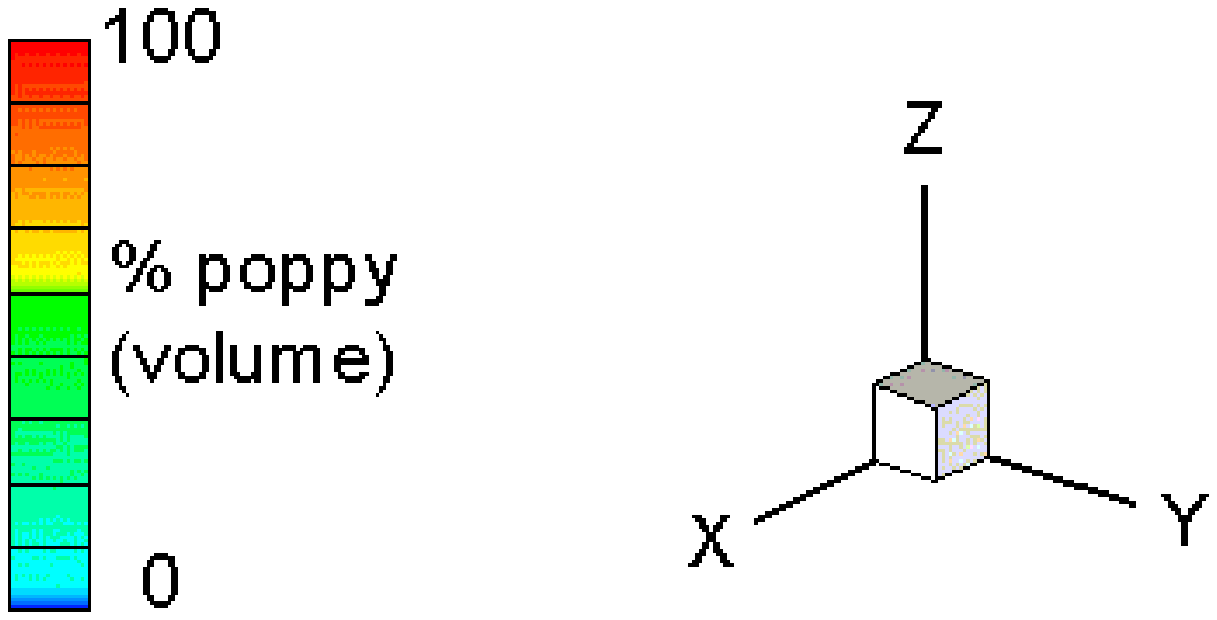}
\caption{\label{fig15} 3D structures showing the progress of one band merging event in a section of a 50\% full and 50 cm long cylinder at (a) 67790 s, (b) 69711 s and (c) 72301 s. The left hand side band is a wall band and therefore stationary during merging.}
\end{figure*}

\begin{figure*}
\includegraphics[width=\columnwidth]{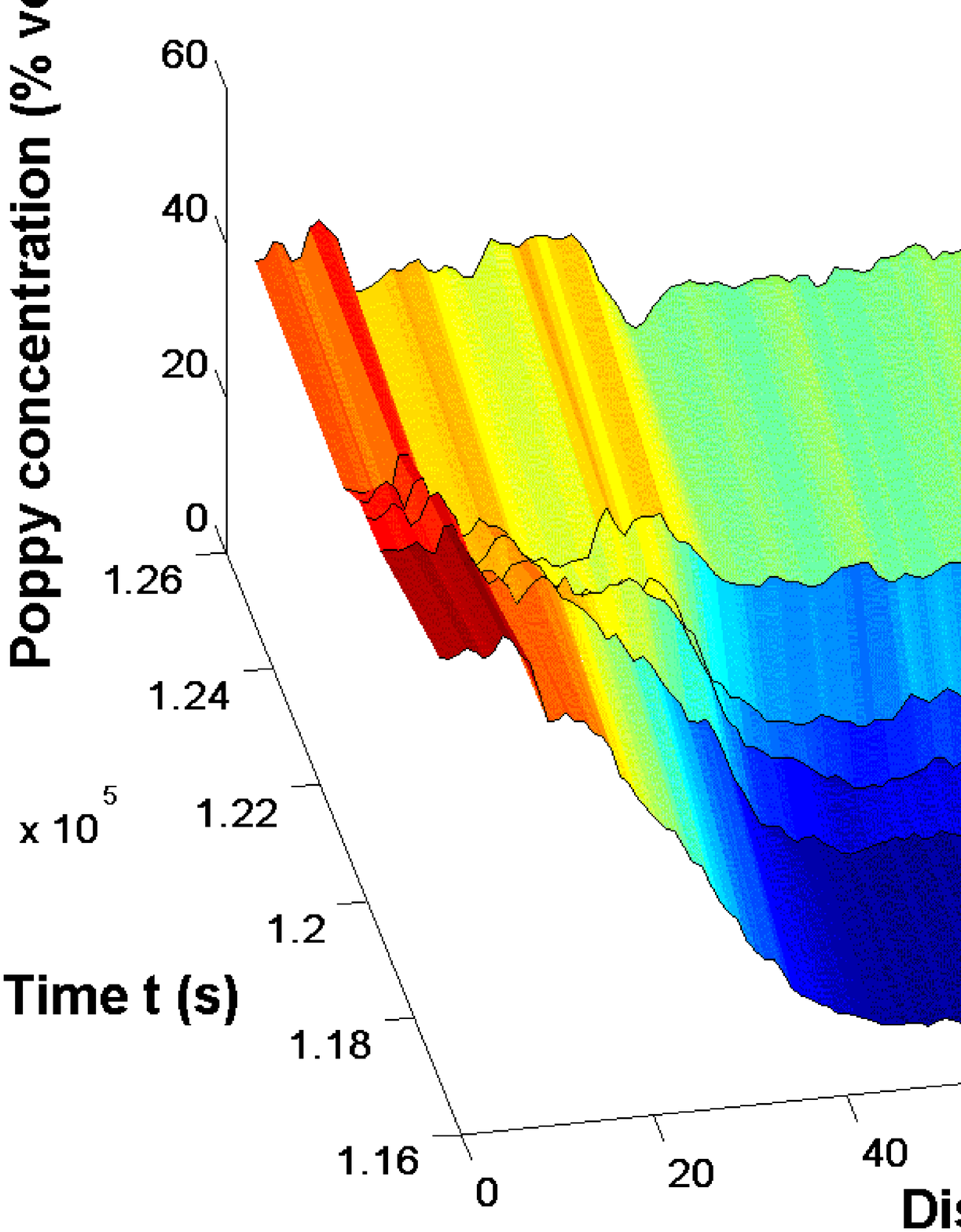}
\hfill \includegraphics[width=0.8\columnwidth]{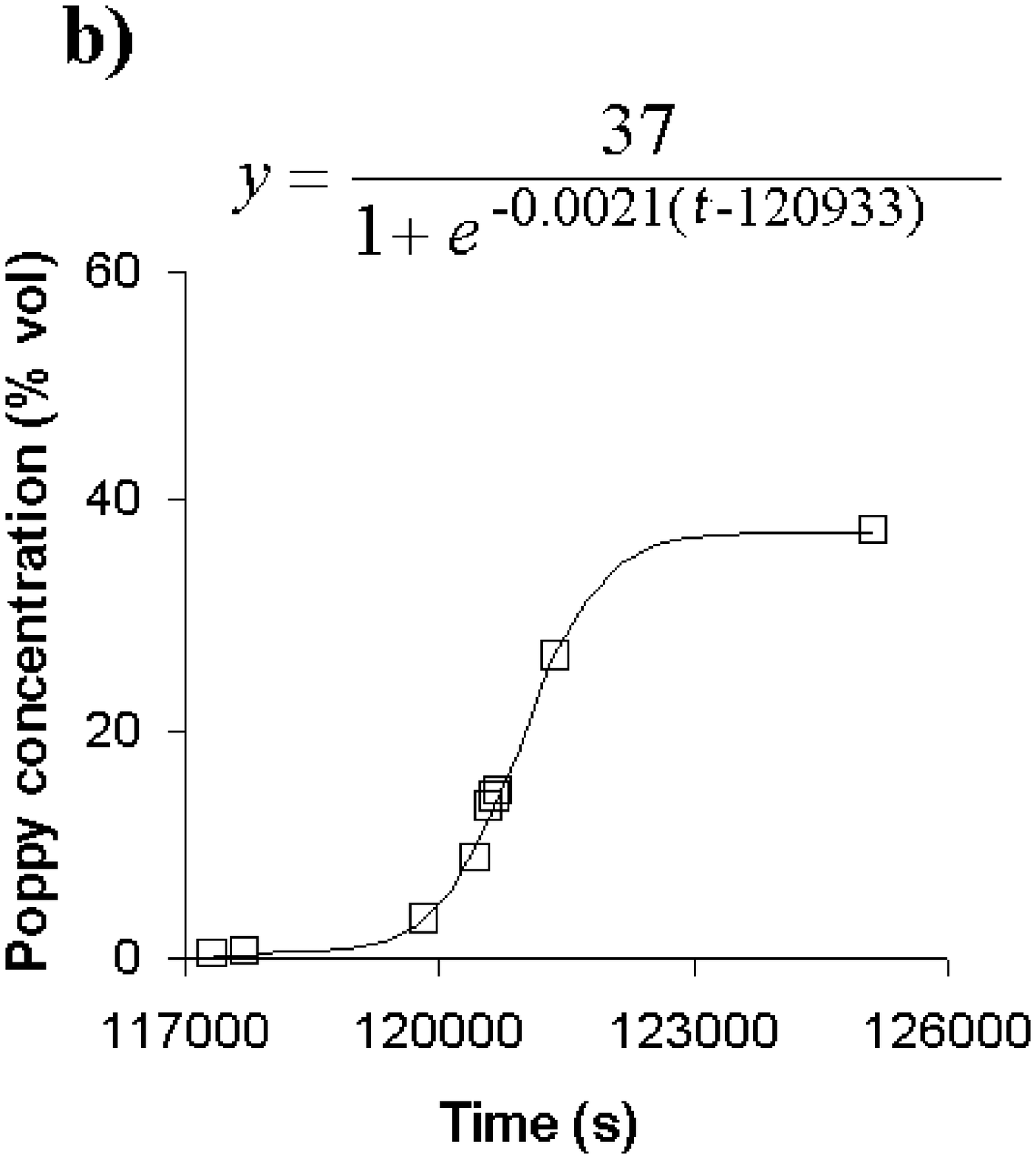}
\caption{\label{fig16} (a) 1D poppy concentration average along the rotation axis of the band merging in Fig.\ \ref{fig15} over time. (b) Poppy concentration of the trough in (a) over time.}
\end{figure*}

First, the images confirm the diffusion of millet seeds through the poppy bands in that the poppy concentration in the poppy bands is slightly reduced during merging as shown in Fig.\ \ref{fig16}(a). This indicates the presence of lower intensity millet seeds inside the poppy bands during the merging process. This is also in agreement with visual inspection of the presence of millets, although mostly on the surface of the poppy bands. 

Second, it is evident that band merging involves the transport of millet seeds from between the poppy bands. Figure \ref{fig16}(a) shows the concentration average along the cylinder length of the last merging event, in which the band on the right hand side is an active band and the other is a stationary band at one end of the cylinder. After 11,700 s (3.25 hours) of rotation, these bands start to merge. The trough between the bands is initially pure millet seeds, i.e., 0\% poppy. The subsequent movement of millet seeds out of the area between the two merging poppy bands is illustrated by the increase of poppy concentration in this region until it equates that in the poppy bands themselves. At this stage, there is still a considerable amount of millets seeds --- 37\% by volume. Millet seeds then keep moving outward, leaving the merged poppy band more concentrated until it reaches the concentration of a fully developed band.

Third, the rate at which band merging occurs can be derived from the acquired images, as shown in Fig.\ \ref{fig16}(b), which shows the concentration minima in the area between the two merging poppy bands over time. It is worth noting that this graph primarily focuses on the time interval in which fast merging occurs. Therefore, it does not show sluggish processes of band traveling prior to merging not the millet removal process in which the poppy concentration gradually increases after merging. A logistic trend is recognized, reinforcing our claim that band traveling occurs very gradually but once merging starts, it takes place very rapidly. Similar graphs, though not shown here, can be generated with two other merging events prior to the one examined above. More abrupt logistics curves are obtained, which means merging occurs more rapidly. This could be attributed to the fact that the second and third last merging events involve two active bands, meaning bands away from the end cap. Therefore millet seeds are free to move through both poppy bands during the merging process hence speeds up merging. However, the last merge is between a fixed band at the end cap and an active poppy band so that millet seeds in between the bands have only one option of traveling through the one active band, thus slows down the millet transport process.

It should also be noted that the ``final'' poppy bands are in fact not pure poppy seeds but contain millet seeds. By MRI, we can quantify this accurately to be 60\% poppy by volume. As mentioned, the newly merged poppy band still contains considerably amount of millet seeds, i.e., 37\% poppy by volume, but becomes more concentrated subject to further rotations until it reaches 60\%, though not shown in Fig.\ \ref{fig16}(b). A confirmation of this figure is by visual inspection and using the length of the final poppy band to compare with the known volume of seeds in the system. The same poppy concentration is found. Also, the same concentration is calculated for all intermediate poppy bands over the course of 80,000 revolutions (38 hours). This is, therefore, considered to be the ultimate poppy concentration. To our knowledge, only visual inspection has been reported that there are some millet seeds in the poppy bands but no quantification is given due to the lack of appropriate tools. This is the first time the poppy concentration inside poppy bands is quantified accurately and non-intrusively by MRI.

\begin{figure*}
\includegraphics[width=0.8\columnwidth]{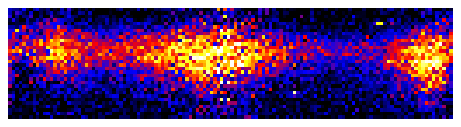}
\includegraphics[width=0.8\columnwidth]{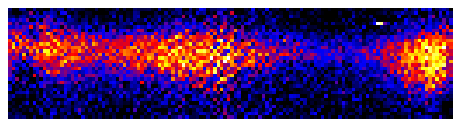}
\includegraphics[width=0.8\columnwidth]{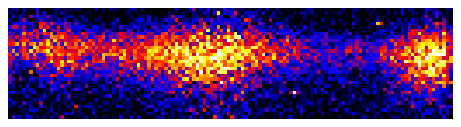}
\includegraphics[width=0.8\columnwidth]{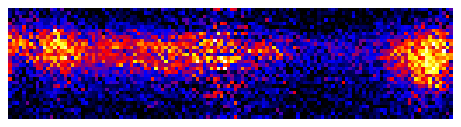}
\includegraphics[width=0.8\columnwidth]{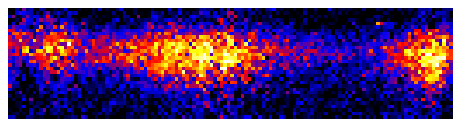}
\includegraphics[width=0.8\columnwidth]{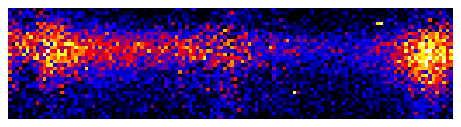}
\includegraphics[width=0.8\columnwidth]{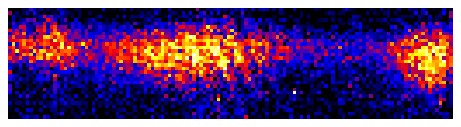}
\includegraphics[width=0.8\columnwidth]{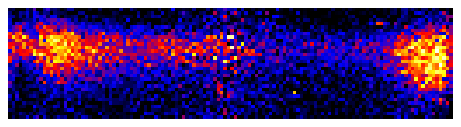}
\includegraphics[width=0.8\columnwidth]{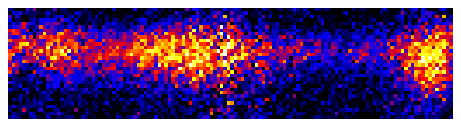}
\includegraphics[width=0.8\columnwidth]{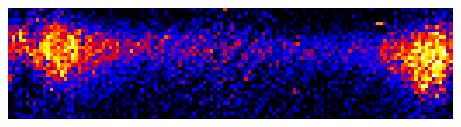}
\caption{\label{fig17} Axial band traveling and merging in a 50\% full and 12 cm long cylinder, rotated at 35 rpm during 30 minute after axial band formation. The timings of these snap shots from the top to bottom, left to right are: 4.7, 5.3, 6.6, 7.1, 7.7, 8.6, 10.5, 11.7, 12.2 and 13.5 min. Color codes: blue: millet, red: mixture, yellow-white: poppy.}
\end{figure*}

As mentioned previously, band traveling and band merging are also observed on a much shorter time scale. A cylinder of 12 cm in length is chosen, in which three temporary poppy bands are formed at full axial segregation. This length is short enough to be fitted within the FOV of the magnet but also long enough to allow three poppy bands to form and merge eventually. The middle band then travels toward and merges with one band at an end cap. For the first time, this dynamic process is imaged in real time by 2D FLASH movie as shown at the provided link and Fig.\ \ref{fig17} capturing 13.5 min of rotation. \\[2mm]

Band traveling and merging: 12 cm long and 50\% full cylinder: \cite{movie10}

\section{\label{sec:level1-4}Conclusions}
In this paper, we present full 3D MRI data of radial and axial segregation in rotating drums. 2D FLASH real time movies are also recorded showing the mechanism of initial radial segregation, the formation of axial bands as well as band traveling and merging. These data should be very useful for modeling work as all spatial information at all intermediate stages of radial and axial segregation are captured by MRI.

Some new observations are made in this work. First, during radial segregation, there is an inner core formed along the rotation axis, which eventually disappears in a 75\% full cylinder but is permanent in an 82\% full cylinder. From this, the avalanche layer thickness can be inferred. Second, band traveling and merging are observed. Prior to each merging event, mostly one poppy band travels at an almost constant speed of 3 $\mu\mbox{m}\ \mbox{s}^{-1}$, until the two merging bands are approximately 3 cm apart. The merging process then involves the sideway traveling in both directions of millet seeds from between the poppy bands until the concentration here equals that of the two poppy bands themselves. Millet seeds then keep traveling out of the merged poppy band, thus shortening this band, until the ultimate poppy concentration of 60\% by volume is reached. The speed of merging follows a logistic rate, which is more rapid if both poppy bands are active, i.e., away from the end cap.  

Initial radial segregation (within the first few rotations) is due to the stopping and percolation of the poppy seeds. Subsequently, the mechanisms of segregation are identified to be both free surface and core diffusion. The latter has largely been ignored in the literature but is shown in this work to become significant in cylinders of more than 75\% full since there exists an inner core buried under the avalanche layer. This core can be clearly seen during radial segregation but becomes axially segregated during axial band formation in an 82\% full cylinder. A future work will follow the current one to study the sub-diffusive nature of this core diffusion phenomenon. 

In addition, it is demonstrated that MRI is a quantitative technique in which the seed concentration at all spatial points can be directly calculated from the signal intensity. The concentration of all poppy bands, including all initial, intermediate and final poppy bands, is quantified accurately for the first time to be 60\% by volume of poppy seeds.

\begin{acknowledgments}
We wish to thank the workshop and the electronic department at the Department of Chemical Engineering for providing experimental equipment and assistance with equipment repair. We would also like to thank Trinity College, Cambridge Overseas Trust, and UK Overseas Research Studentship for financial support. We would like to thank Dr.\ Christian Mikkelsen for many interesting discussions.
\end{acknowledgments}

\bibliography{granular-ref}

\begin{thebibliography}{28}
\expandafter\ifx\csname natexlab\endcsname\relax\def\natexlab#1{#1}\fi
\expandafter\ifx\csname bibnamefont\endcsname\relax
  \def\bibnamefont#1{#1}\fi
\expandafter\ifx\csname bibfnamefont\endcsname\relax
  \def\bibfnamefont#1{#1}\fi
\expandafter\ifx\csname citenamefont\endcsname\relax
  \def\citenamefont#1{#1}\fi
\expandafter\ifx\csname url\endcsname\relax
  \def\url#1{\texttt{#1}}\fi
\expandafter\ifx\csname urlprefix\endcsname\relax\def\urlprefix{URL }\fi
\providecommand{\bibinfo}[2]{#2}
\providecommand{\eprint}[2][]{\url{#2}}

\bibitem[{\citenamefont{Painter et~al.}(1997)\citenamefont{Painter, Tennakoon,
  and Behringer}}]{painter97}
\bibinfo{author}{\bibfnamefont{B.}~\bibnamefont{Painter}},
  \bibinfo{author}{\bibfnamefont{S.}~\bibnamefont{Tennakoon}},
  \bibnamefont{and} \bibinfo{author}{\bibfnamefont{R.~P.}
  \bibnamefont{Behringer}}, in \emph{\bibinfo{booktitle}{Physics of dry
  granular media}}, edited by \bibinfo{editor}{\bibfnamefont{H.~J.}
  \bibnamefont{Herrmann}}, \bibinfo{editor}{\bibfnamefont{J.~P.}
  \bibnamefont{Hovi}}, \bibnamefont{and}
  \bibinfo{editor}{\bibfnamefont{S.}~\bibnamefont{Luding}}
  (\bibinfo{publisher}{Kluwer Academic Publishers}, \bibinfo{year}{1997}), p.
  \bibinfo{pages}{217}.

\bibitem[{\citenamefont{Gioia et~al.}(2006)\citenamefont{Gioia, Ott-Monsivais,
  and Hill}}]{gioia06}
\bibinfo{author}{\bibfnamefont{G.}~\bibnamefont{Gioia}},
  \bibinfo{author}{\bibfnamefont{S.~E.} \bibnamefont{Ott-Monsivais}},
  \bibnamefont{and} \bibinfo{author}{\bibfnamefont{K.~M.} \bibnamefont{Hill}},
  \bibinfo{journal}{Phys.\ Rev.\ Lett.} \textbf{\bibinfo{volume}{96}},
  \bibinfo{pages}{138001} (\bibinfo{year}{2006}).

\bibitem[{\citenamefont{Rajchenbach}(1997)}]{rajchenbach97}
\bibinfo{author}{\bibfnamefont{J.}~\bibnamefont{Rajchenbach}}, in
  \emph{\bibinfo{booktitle}{Physics of dry granular media}}, edited by
  \bibinfo{editor}{\bibfnamefont{H.~J.} \bibnamefont{Herrmann}},
  \bibinfo{editor}{\bibfnamefont{J.~P.} \bibnamefont{Hovi}}, \bibnamefont{and}
  \bibinfo{editor}{\bibfnamefont{S.}~\bibnamefont{Luding}}
  (\bibinfo{publisher}{Kluwer Academic Publishers}, \bibinfo{year}{1997}), p.
  \bibinfo{pages}{421}.

\bibitem[{\citenamefont{Ottino and Khakhar}(2000)}]{ottino00}
\bibinfo{author}{\bibfnamefont{J.~M.} \bibnamefont{Ottino}} \bibnamefont{and}
  \bibinfo{author}{\bibfnamefont{D.~V.} \bibnamefont{Khakhar}},
  \bibinfo{journal}{Annu.\ Rev.\ Fluid\ Mech.} \textbf{\bibinfo{volume}{32}},
  \bibinfo{pages}{55} (\bibinfo{year}{2000}).

\bibitem[{\citenamefont{Hill et~al.}(2005)\citenamefont{Hill, Gioia, Amaravadi,
  and Winter}}]{hill05}
\bibinfo{author}{\bibfnamefont{K.~M.} \bibnamefont{Hill}},
  \bibinfo{author}{\bibfnamefont{G.}~\bibnamefont{Gioia}},
  \bibinfo{author}{\bibfnamefont{D.}~\bibnamefont{Amaravadi}},
  \bibnamefont{and} \bibinfo{author}{\bibfnamefont{C.}~\bibnamefont{Winter}},
  \bibinfo{journal}{Complexity} \textbf{\bibinfo{volume}{10}},
  \bibinfo{pages}{79} (\bibinfo{year}{2005}).

\bibitem[{\citenamefont{Clement et~al.}(1995)\citenamefont{Clement,
  Rajchenbach, and Duran}}]{clement95}
\bibinfo{author}{\bibfnamefont{E.}~\bibnamefont{Clement}},
  \bibinfo{author}{\bibfnamefont{J.}~\bibnamefont{Rajchenbach}},
  \bibnamefont{and} \bibinfo{author}{\bibfnamefont{J.}~\bibnamefont{Duran}},
  \bibinfo{journal}{Europhys.\ Lett.} \textbf{\bibinfo{volume}{30}},
  \bibinfo{pages}{7} (\bibinfo{year}{1995}).

\bibitem[{\citenamefont{Edwards and Oakshott}(1989)}]{edwards89}
\bibinfo{author}{\bibfnamefont{S.~F.} \bibnamefont{Edwards}} \bibnamefont{and}
  \bibinfo{author}{\bibfnamefont{R.~B.~S.} \bibnamefont{Oakshott}},
  \bibinfo{journal}{Physica A} \textbf{\bibinfo{volume}{157}},
  \bibinfo{pages}{1080} (\bibinfo{year}{1989}).

\bibitem[{\citenamefont{Cantelaube and Bideau}(1995)}]{cantelaube95}
\bibinfo{author}{\bibfnamefont{F.}~\bibnamefont{Cantelaube}} \bibnamefont{and}
  \bibinfo{author}{\bibfnamefont{D.}~\bibnamefont{Bideau}},
  \bibinfo{journal}{Europhys.\ Lett.} \textbf{\bibinfo{volume}{30}},
  \bibinfo{pages}{133} (\bibinfo{year}{1995}).

\bibitem[{\citenamefont{Lai et~al.}(1997)\citenamefont{Lai, Jia, and
  Chan}}]{lai97}
\bibinfo{author}{\bibfnamefont{P.-Y.} \bibnamefont{Lai}},
  \bibinfo{author}{\bibfnamefont{L.-C.} \bibnamefont{Jia}}, \bibnamefont{and}
  \bibinfo{author}{\bibfnamefont{C.~K.} \bibnamefont{Chan}},
  \bibinfo{journal}{Phys.\ Rev.\ Lett.} \textbf{\bibinfo{volume}{79}},
  \bibinfo{pages}{4994} (\bibinfo{year}{1997}).

\bibitem[{\citenamefont{Khakhar et~al.}(1997)\citenamefont{Khakhar, McCarthy,
  and Ottino}}]{khakhar97}
\bibinfo{author}{\bibfnamefont{D.~V.} \bibnamefont{Khakhar}},
  \bibinfo{author}{\bibfnamefont{J.~J.} \bibnamefont{McCarthy}},
  \bibnamefont{and} \bibinfo{author}{\bibfnamefont{J.~M.}
  \bibnamefont{Ottino}}, \bibinfo{journal}{Phys.\ Fluids}
  \textbf{\bibinfo{volume}{9}}, \bibinfo{pages}{3600} (\bibinfo{year}{1997}).

\bibitem[{\citenamefont{McCarthy et~al.}(1996)\citenamefont{McCarthy, Shinbrot,
  Metcalfe, Wolfe, and Ottino}}]{mccarthy96}
\bibinfo{author}{\bibfnamefont{J.~J.} \bibnamefont{McCarthy}},
  \bibinfo{author}{\bibfnamefont{T.}~\bibnamefont{Shinbrot}},
  \bibinfo{author}{\bibfnamefont{G.}~\bibnamefont{Metcalfe}},
  \bibinfo{author}{\bibfnamefont{J.~E.} \bibnamefont{Wolfe}}, \bibnamefont{and}
  \bibinfo{author}{\bibfnamefont{J.~M.} \bibnamefont{Ottino}},
  \bibinfo{journal}{AIChE} \textbf{\bibinfo{volume}{42}}, \bibinfo{pages}{3351}
  (\bibinfo{year}{1996}).

\bibitem[{\citenamefont{Hill et~al.}(1997)\citenamefont{Hill, Caprihan, and
  Kakalios}}]{hill97}
\bibinfo{author}{\bibfnamefont{K.~M.} \bibnamefont{Hill}},
  \bibinfo{author}{\bibfnamefont{A.}~\bibnamefont{Caprihan}}, \bibnamefont{and}
  \bibinfo{author}{\bibfnamefont{J.}~\bibnamefont{Kakalios}},
  \bibinfo{journal}{Phys.\ Rev.\ E} \textbf{\bibinfo{volume}{56}},
  \bibinfo{pages}{4386} (\bibinfo{year}{1997}).

\bibitem[{\citenamefont{Newey et~al.}(2004)\citenamefont{Newey, Ozik, Meer,
  Ott, and Losert}}]{newey04}
\bibinfo{author}{\bibfnamefont{M.}~\bibnamefont{Newey}},
  \bibinfo{author}{\bibfnamefont{J.}~\bibnamefont{Ozik}},
  \bibinfo{author}{\bibfnamefont{S.~M. V.~D.} \bibnamefont{Meer}},
  \bibinfo{author}{\bibfnamefont{E.}~\bibnamefont{Ott}}, \bibnamefont{and}
  \bibinfo{author}{\bibfnamefont{W.}~\bibnamefont{Losert}},
  \bibinfo{journal}{Europhys.\ Lett.} \textbf{\bibinfo{volume}{66}},
  \bibinfo{pages}{205} (\bibinfo{year}{2004}).

\bibitem[{\citenamefont{Choo et~al.}(1998)\citenamefont{Choo, Baker, Molteno,
  and Morris}}]{choo98}
\bibinfo{author}{\bibfnamefont{K.}~\bibnamefont{Choo}},
  \bibinfo{author}{\bibfnamefont{M.~W.} \bibnamefont{Baker}},
  \bibinfo{author}{\bibfnamefont{T.~C.~A.} \bibnamefont{Molteno}},
  \bibnamefont{and} \bibinfo{author}{\bibfnamefont{S.~W.}
  \bibnamefont{Morris}}, \bibinfo{journal}{Phys.\ Rev.\ E}
  \textbf{\bibinfo{volume}{58}}, \bibinfo{pages}{6115} (\bibinfo{year}{1998}).

\bibitem[{\citenamefont{Aranson and Tsimring}(1999)}]{aranson99}
\bibinfo{author}{\bibfnamefont{I.}~\bibnamefont{Aranson}} \bibnamefont{and}
  \bibinfo{author}{\bibfnamefont{L.~S.} \bibnamefont{Tsimring}},
  \bibinfo{journal}{Phys.\ Rev.\ Lett.} \textbf{\bibinfo{volume}{82}},
  \bibinfo{pages}{4643} (\bibinfo{year}{1999}).

\bibitem[{\citenamefont{Fiedor and Ottino}(2003)}]{fiedor03}
\bibinfo{author}{\bibfnamefont{S.~J.} \bibnamefont{Fiedor}} \bibnamefont{and}
  \bibinfo{author}{\bibfnamefont{J.~M.} \bibnamefont{Ottino}},
  \bibinfo{journal}{Phys.\ Rev.\ Lett.} \textbf{\bibinfo{volume}{91}},
  \bibinfo{pages}{244301} (\bibinfo{year}{2003}).

\bibitem[{\citenamefont{Callaghan}(1991)}]{callaghan91}
\bibinfo{author}{\bibfnamefont{P.~T.} \bibnamefont{Callaghan}},
  \emph{\bibinfo{title}{Principles of Nuclear Magnetic Resonance Microscopy}}
  (\bibinfo{publisher}{Oxford Science Publications}, \bibinfo{year}{1991}).

\bibitem[{\citenamefont{Hennig et~al.}(1986)\citenamefont{Hennig, Nauerth, and
  Friedburg}}]{hennig86}
\bibinfo{author}{\bibfnamefont{J.}~\bibnamefont{Hennig}},
  \bibinfo{author}{\bibfnamefont{A.}~\bibnamefont{Nauerth}}, \bibnamefont{and}
  \bibinfo{author}{\bibfnamefont{H.}~\bibnamefont{Friedburg}},
  \bibinfo{journal}{Magn.\ Res.\ Med.} \textbf{\bibinfo{volume}{3}},
  \bibinfo{pages}{823} (\bibinfo{year}{1986}).

\bibitem[{\citenamefont{Haase et~al.}(1986)\citenamefont{Haase, Frahm,
  Matthaei, Hanicke, and Merboldt}}]{haase86}
\bibinfo{author}{\bibfnamefont{A.}~\bibnamefont{Haase}},
  \bibinfo{author}{\bibfnamefont{J.}~\bibnamefont{Frahm}},
  \bibinfo{author}{\bibfnamefont{S.}~\bibnamefont{Matthaei}},
  \bibinfo{author}{\bibfnamefont{W.}~\bibnamefont{Hanicke}}, \bibnamefont{and}
  \bibinfo{author}{\bibfnamefont{K.~D.} \bibnamefont{Merboldt}},
  \bibinfo{journal}{J.\ Magn.\ Reson.} \textbf{\bibinfo{volume}{67}},
  \bibinfo{pages}{258} (\bibinfo{year}{1986}).

\bibitem[{\citenamefont{Mantle and Sederman}(2003)}]{mantle03}
\bibinfo{author}{\bibfnamefont{M.~D.} \bibnamefont{Mantle}} \bibnamefont{and}
  \bibinfo{author}{\bibfnamefont{A.~J.} \bibnamefont{Sederman}},
  \bibinfo{journal}{Prog.\ Nuc.\ Magn.\ Reson.\ Spec.}
  \textbf{\bibinfo{volume}{43}}, \bibinfo{pages}{3} (\bibinfo{year}{2003}).

\bibitem[{\citenamefont{Kakalios}(2005)}]{kakalios05}
\bibinfo{author}{\bibfnamefont{J.}~\bibnamefont{Kakalios}},
  \bibinfo{journal}{Am.\ J.\ Phys.} \textbf{\bibinfo{volume}{73}},
  \bibinfo{pages}{8} (\bibinfo{year}{2005}).

\bibitem[{\citenamefont{Jaeger and Nagel}(1992)}]{jaeger92}
\bibinfo{author}{\bibfnamefont{H.~M.} \bibnamefont{Jaeger}} \bibnamefont{and}
  \bibinfo{author}{\bibfnamefont{S.~R.} \bibnamefont{Nagel}},
  \bibinfo{journal}{Science} \textbf{\bibinfo{volume}{256}},
  \bibinfo{pages}{1523} (\bibinfo{year}{1992}).

\bibitem[{\citenamefont{Seymour et~al.}(2000)\citenamefont{Seymour, Caprihan,
  Altobelli, and Fukushima}}]{seymour00}
\bibinfo{author}{\bibfnamefont{J.~D.} \bibnamefont{Seymour}},
  \bibinfo{author}{\bibfnamefont{A.}~\bibnamefont{Caprihan}},
  \bibinfo{author}{\bibfnamefont{S.~A.} \bibnamefont{Altobelli}},
  \bibnamefont{and}
  \bibinfo{author}{\bibfnamefont{E.}~\bibnamefont{Fukushima}},
  \bibinfo{journal}{Phys.\ Rev.\ Lett.} \textbf{\bibinfo{volume}{84}},
  \bibinfo{pages}{266} (\bibinfo{year}{2000}).

\bibitem[{\citenamefont{Yamane et~al.}(1998)\citenamefont{Yamane, Nakagama,
  Altobelli, Tanaka, and Tsuji}}]{yamane98}
\bibinfo{author}{\bibfnamefont{K.}~\bibnamefont{Yamane}},
  \bibinfo{author}{\bibfnamefont{M.}~\bibnamefont{Nakagama}},
  \bibinfo{author}{\bibfnamefont{S.~A.} \bibnamefont{Altobelli}},
  \bibinfo{author}{\bibfnamefont{T.}~\bibnamefont{Tanaka}}, \bibnamefont{and}
  \bibinfo{author}{\bibfnamefont{Y.}~\bibnamefont{Tsuji}},
  \bibinfo{journal}{Phys.\ Fluids} \textbf{\bibinfo{volume}{10}},
  \bibinfo{pages}{1419} (\bibinfo{year}{1998}).

\bibitem[{mov({\natexlab{a}})}]{movie1}
\eprint{See EPAPS Document No. number will be inserted by publisher for
  ``Movie82-10rpm64.mpg'': Radial segregation, 82\% full cylinder at 10 rpm}.

\bibitem[{\citenamefont{Zik et~al.}(1994)\citenamefont{Zik, Levine, Lipson,
  Shtrikman, and Stavans}}]{zik94}
\bibinfo{author}{\bibfnamefont{O.}~\bibnamefont{Zik}},
  \bibinfo{author}{\bibfnamefont{D.}~\bibnamefont{Levine}},
  \bibinfo{author}{\bibfnamefont{S.~G.} \bibnamefont{Lipson}},
  \bibinfo{author}{\bibfnamefont{S.}~\bibnamefont{Shtrikman}},
  \bibnamefont{and} \bibinfo{author}{\bibfnamefont{J.}~\bibnamefont{Stavans}},
  \bibinfo{journal}{Phys.\ Rev.\ Lett.} \textbf{\bibinfo{volume}{73}},
  \bibinfo{pages}{644} (\bibinfo{year}{1994}).

\bibitem[{mov({\natexlab{b}})}]{movie9}
\eprint{See EPAPS Document No. number will be inserted by publisher for
  ``Movie50-12cmft-form.mpg'': Axial band formation, 50\% full, 12 cm
  cylinder}.

\bibitem[{mov({\natexlab{c}})}]{movie10}
\eprint{See EPAPS Document No. number will be inserted by publisher for
  ``Movie50-12cmft-merge.mpg'': Band traveling and merging, 50\% full, 12 cm
  cylinder}.

\end{thebibliography}

\end{document}